\documentclass[acmtog,nonacm, screen]{acmart}
\AtBeginDocument{%
  }

\usepackage{algpseudocode}
\usepackage{algorithm}
\usepackage{tikz}
\usepackage{colortbl}
\usepackage{subcaption}
\usepackage{microtype}
\usepackage{multirow}
\usepackage{booktabs}
\usepackage{placeins}
\usepackage{pifont}
\usepackage{wrapfig}
\usepackage{todonotes}
\usepackage{xspace}
\usepackage{enumitem}
\usepackage[table,usenames,dvipsnames]{xcolor}

\newcommand{\appref}[1]{Suppl. \ref*{#1}\xspace}
\newcommand{\ie}{\textit{i.e.}\xspace}
\newcommand{\eg}{\textit{e.g.}\xspace}

\newcommand{\cmark}{\textcolor{ForestGreen}{\ding{51}}}%
\newcommand{\xmark}{\textcolor{BrickRed}{\ding{55}}}%
\setcopyright{acmlicensed}
\copyrightyear{2018}
\acmYear{2018}
\acmDOI{XXXXXXX.XXXXXXX}
\acmConference[SIGGRAPH '26]{Make sure to enter the correct
  conference title from your rights confirmation email}{June 03--05,
  2018}{Woodstock, NY}
\acmISBN{978-1-4503-XXXX-X/2018/06}

\acmSubmissionID{395}

\begin{document}

\title{A LoD of Gaussians: Out-of-Core Training and Rendering for Seamless Ultra-Large Scene Reconstruction}

\author{Felix Windisch}
\email{felix.windisch@tugraz.at}
\orcid{1234-5678-9012}
\affiliation{%
	\institution{Graz University of Technology}
	\city{Graz}
	\country{Austria}
}

\author{Thomas K\"ohler}
\affiliation{%
	\institution{Graz University of Technology}
	\country{Austria}}
\email{t.koehler@tugraz.at}

\author{Lukas Radl}
\affiliation{%
	\institution{Graz University of Technology}
	\country{Austria}}
\email{lukas.radl@tugraz.at}

\author{Mattia D'Urso}
\email{	mattia.durso@tugraz.at}
\affiliation{%
	\institution{Graz University of Technology}
	\country{Austria}}

\author{Michael Steiner}
\email{michael.steiner@tugraz.at}
\affiliation{%
	\institution{Graz University of Technology}
	\country{Austria}}

\author{Dieter Schmalstieg}
\affiliation{%
	\institution{Graz University of Technology}
	\country{Austria}
}
\affiliation{%
	\institution{University of Stuttgart}
	\country{Germany}}
\email{dieter.schmalstieg@visus.uni-stuttgart.de}

\author{Markus Steinberger}
\affiliation{%
	\institution{Graz University of Technology}
	\country{Austria}
}
\affiliation{%
	\institution{Huawei Technologies}
	\country{Austria}
}


\begin{abstract}
 Gaussian Splatting has emerged as a high-performance technique for novel view synthesis, enabling real-time rendering and high-quality reconstruction of small scenes. However, scaling to larger environments has so far relied on partitioning the scene into chunks---a strategy that introduces artifacts at chunk boundaries, complicates training across varying scales, and is poorly suited to unstructured scenarios such as city-scale flyovers combined with street-level views. Moreover, rendering remains fundamentally limited by GPU memory, as all visible chunks must reside in VRAM simultaneously.
We introduce \emph{A LoD of Gaussians}, a framework for training and rendering ultra-large-scale Gaussian scenes on a single consumer-grade GPU without partitioning. Our method stores the full scene out-of-core (\eg, in CPU memory) and trains a Level-of-Detail (LoD) representation directly, dynamically streaming only the relevant Gaussians. A hybrid data structure combining Gaussian hierarchies with Sequential Point Trees enables efficient, view-dependent LoD selection, while a lightweight caching and view scheduling system exploits temporal coherence to minimize the loading overhead. Together, these innovations enable seamless multi-scale reconstruction and interactive visualization of complex scenes---from broad aerial views to fine-grained ground-level details.
\end{abstract}


\begin{teaserfigure}
  \includegraphics[width=\textwidth]{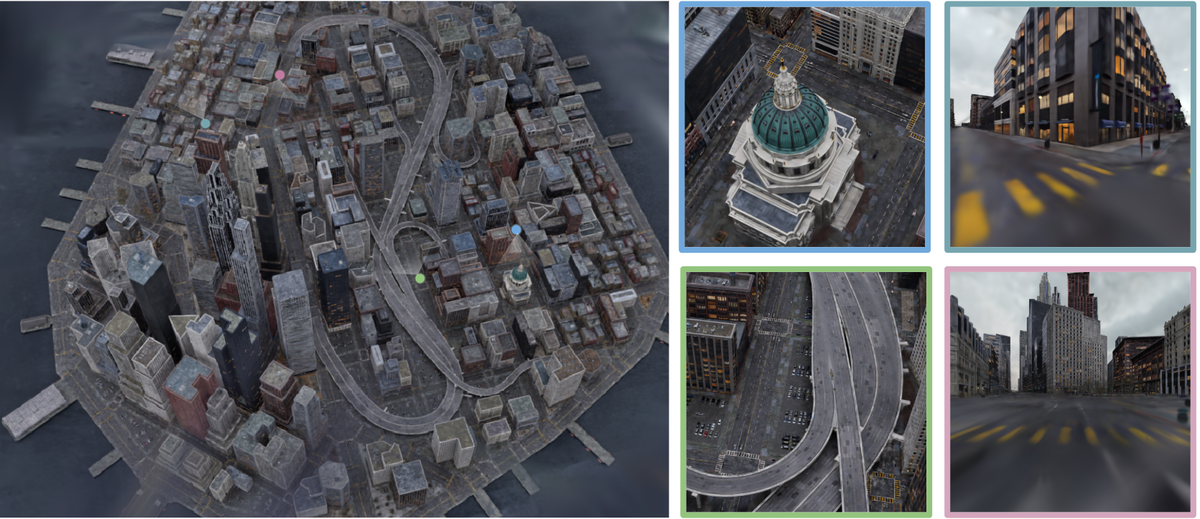}
  \caption{
\textbf{Teaser}: 
We introduce a fully hierarchical 3D Gaussian representation trained directly across unstructured, multi-scale image sets—including street-level and far aerial views—without scene partitioning. Our method maintains a consistent global scene model, eliminating boundary artifacts typical of chunked approaches. A hybrid Level-of-Detail system combines Gaussian hierarchies with Sequential Point Trees, enabling dynamic, view-dependent streaming and LoD selection. The entire model resides in external memory, with only a small, adaptive subset loaded on demand, allowing seamless training and interactive rendering of scenes with 150M+ Gaussians on a single consumer GPU ($\leq24GB$ VRAM). }
  \label{fig:teaser}
\end{teaserfigure}


\maketitle

\section{Introduction}
Given a set of posed images of a 3D scene, the task of novel view synthesis (NVS) is to generate plausible images of the scene from unseen viewpoints.
Early approaches achieved this via image-based blending \citep{DeepBlending}, but the introduction of Neural Radiance Fields (NeRF) \citep{NeRF} marked a breakthrough, enabling high-quality results by optimizing an implicit volumetric scene representation through a multi-layer perceptron.
More recently, 3D Gaussian Splatting (3DGS) \citep{3DGS} extended this paradigm to an explicit representation: a set of Gaussian primitives that are efficiently rasterized using splatting techniques \citep{zwicker2001EWA}, replacing costly ray marching and allowing real-time rendering with fast convergence.

Despite these advances, both NeRF and 3DGS remain constrained by memory bottlenecks when applied to large-scale environments. Prior methods address this by dividing scenes into smaller chunks \citep{GridNeRF, BlockNeRF, Hierarchical, CityGaussian, GigaGS, VastGaussian}, training each independently before merging results. While chunking strategies mitigate memory usage during training, they introduce several key limitations:
%
\begin{enumerate}[leftmargin=14pt]
    \item \textbf{View-chunk misalignment:} Camera views often span multiple chunks, especially in open or multi-scale datasets (\eg, combining aerial and street-level images). As a result, chunk boundaries become arbitrary with respect to images, complicating scene partitioning and training, resulting in artifacts: \textit{Chunk bleeding} occurs when Gaussians extend out of their assigned chunk and obscure neighbouring chunks after merging. \textit{Chunk ghosting} occurs when an occluder present in a training image is not part of the current chunk, training its `ghostly' outline into the wrong chunk. See Figures \ref{fig:artifacts} and \ref{fig:Quality}.
    \item \textbf{Redundant overlap: } To avoid artifacts at chunk boundaries, regions are typically trained with significant overlap, which duplicates parameters and optimizer state, increasing memory usage and prolonging training time.
    \item \textbf{Asymmetric hardware demands: } Although chunking reduces memory requirements during training, rendering may require all visible chunks in memory simultaneously---often exceeding the capacity of the original training setup and undermining the practical benefit of partitioning. 
\end{enumerate}

The simplest and most robust alternative to chunking is to avoid splitting altogether.
With \emph{A LoD of Gaussians}, we introduce a seamless pipeline that enables training and rendering of ultra-large-scale scenes directly on a single consumer-grade GPU, without any form of scene partitioning (see Figure \ref{fig:Training}). To handle scenes that exceed available VRAM, we store all Gaussian data in CPU RAM and dynamically stream only those visible from the current training view into GPU memory. Still, a single distant view could require access to the full scene. To address this, we construct a hierarchical Level-of-Detail (LoD) model inspired by \citet{Hierarchical}, loading detail proportional to view distance. Crucially, this hierarchy must be maintained throughout training, as Gaussian parameters and their spatial distribution evolve dynamically. We propose a novel hierarchy densification strategy, inspired by MCMC-style spawning \citep{3DGSasMCMC}, to support stable and progressive refinement.
Efficient view-dependent selection from the hierarchy is challenging for large models, particularly when the hierarchy structure changes during optimization. Instead of full tree traversal, we adopt Sequential Point Trees (SPTs) \citep{SPT}, originally developed for point cloud rendering. Our Hierarchical SPT version allows us to compute the correct LoD cut efficiently for  individual views and camera paths.
Finally, to reduce CPU-GPU data transfer, we introduce a lightweight caching system that tracks recently used Gaussians and reuses them across training iterations.
In summary, we make the following contributions:
\begin{enumerate}[leftmargin=14pt]
\item \textbf{Seamless, non-partitioned training of ultra-large Gaussian scenes.}
We present the first 3D Gaussian Splatting framework that enables training and interactive rendering of city-scale scenes from arbitrary views without spatial partitioning, using out-of-core memory and view-dependent streaming on a single consumer GPU.
\item \textbf{Dynamic LoD hierarchy densification during training.}
We propose a novel coarse-to-fine hierarchy densification strategy that allows Gaussian LoD hierarchies to evolve continuously during optimization, supporting stable refinement and restructuring without post-training hierarchy construction.
\item \textbf{Hierarchical Sequential Point Trees for Gaussian Splatting.}
We adapt Sequential Point Trees (SPTs) to Gaussian splatting and introduce the Hierarchical SPT (HSPT), a hybrid data structure  enabling efficient, parallelizable LoD selection while remaining robust to hierarchy updates during training.
\item \textbf{Efficient out-of-core execution via cache-aware streaming and view scheduling.}
We design a lightweight GPU caching and view selection strategy that exploits temporal coherence to substantially reduce CPU–GPU transfer overhead during both training and rendering.
\item \textbf{Large-scale evaluation and new benchmarks.}
We introduce the Uni10k dataset and an expanded version of the MatrixCity Small City scene, and demonstrate state-of-the-art performance on large-scale, multi-view datasets spanning aerial, street-level, and indoor environments.
\end{enumerate}

\section{Related Work}

\paragraph*{Large Scale Reconstruction}
Reconstructing large-scale scenes from images has long been a central challenge in visual computing. 
Traditional approaches relied on Structure-from-Motion (SfM) pipelines to recover geometry from unordered photo collections \citep{RomeInADay, COLMAP}.
Differentiable rendering techniques, notably NeRF \citep{NeRF} and 3DGS \citep{3DGS}, marked a paradigm shift by optimizing volumetric scene representations. Extensions of NeRF to large scenes typically employ scene partitioning \citep{BlockNeRF, GridNeRF} or multi-GPU training strategies \citep{NerfXL}. Similarly, most large-scale 3DGS pipelines adopt chunk-based training: \emph{Hierarchical-3DGS} (\emph{H-3DGS}) \citep{Hierarchical} trains chunks independently and then merges them into a global LoD hierarchy; \emph{CityGaussian} \citep{CityGaussian} combines chunked training with per-chunk LoD selection using \emph{LightGaussian} \citep{LightGaussian}; \emph{OccluGaussian} \citep{OccluGaussian} partitions the scene to maximize camera correlation in each chunk; and \emph{VastGaussian} \citep{VastGaussian} introduces decoupled appearance modeling and progressive partitioning. \emph{Horizon-GS} \citep{HorizonGS} integrates divide-and-conquer strategies with LoD mechanisms from \citet{OctreeGS}, specifically targeting hybrid aerial/street-view datasets. \emph{GrendelGS} \citep{Grendel} avoids spatial chunking by distributing training images across GPUs, such that each device renders a disjoint screen region.
Another research direction focuses on extracting geometric proxies from large-scale 3DGS scenes \citep{ULSRGS, CityGaussianV2, GigaGS}. These methods leverage TSDF fusion and geometric losses to generate multiple meshes, which are fused and rendered efficiently using traditional rasterization.
Other approaches \citep{CLM, GS-Scale} make use out-of-core memory to scale training, but neglect LoD, which limits their memory reduction during training to frustum culling.


\paragraph*{Level-of-Detail Rendering}
Level-of-detail techniques reduce geometric complexity of distant scene content to accelerate rendering. In 3DGS, LoD approaches have mainly targeted efficient rendering on memory-constrained or mobile devices. Compression-based strategies include attribute quantization via codebooks, pruning low-impact Gaussians, and adapting the degree of spherical harmonics per primitive \citep{ReducingTheMemoryFootprint, Compressed3DGaussianSplatting, LightGaussian, MiniSplatting, HierarchicalCompression, RadSplat, FLoD}. \emph{Scaffold-GS} \citep{ScaffoldGS} uses latent vectors anchored to reference Gaussians, with an MLP generating associated Gaussians at render time. \emph{Octree-GS} \citep{OctreeGS} extends this to hierarchical LoD rendering via spatial subdivision. \emph{Virtualized 3D Gaussians} \citep{Virtualized3DGaussians} targets composed scenes of reconstructed objects using an LoD scheme inspired by Unreal Engine 5's \emph{Nanite} \citep{Nanite}.

\section{Preliminaries of Hierarchical 3D Gaussian Splatting}


3DGS \citep{3DGS} models a radiance field using a set of spatially distributed Gaussians, each with mean $\boldsymbol{\mu}_i \in \mathbb{R}^3$, RGB base colors $\mathbf{b}_i \in \mathbb{R}^3$ and covariance matrices $ \boldsymbol{\Sigma}_i = \mathbf{R}_i \mathbf{S}_i \mathbf{S}_i^\top \mathbf{R}_i^\top$, which are parameterized via a diagonal scaling matrix $\mathbf{S}_i = \text{diag}(s^1_i, s^2_i, s^3_i) $ and an orthonormal rotation matrix $\mathbf{R}_i$.
Each Gaussian also stores an opacity $\sigma_i$ and view-dependent color, modeled using spherical harmonics (SH) coefficients $\mathbf{f}_i^d$. The SH degree $d$ controls expressiveness, with each Gaussian requiring $\sum_{j=1}^{d}3 \cdot (2j + 1)$ parameters.
For rendering, all $N$ Gaussians are sorted by distance to the camera and a discrete approximation of the volume rendering equation is evaluated for every pixel $\mathbf{x}$ with corresponding view direction $\mathbf{v}$:
\begin{equation}
    \mathbf{C}(\mathbf{x}) = \sum_{i=1}^N \mathbf{c}_i(\mathbf{v}) \alpha_i(\mathbf{x}) \prod_{j=1}^{i-1} (1 - \alpha_j(\mathbf{x})) ,\end{equation}
where $\alpha_j$ is the opacity of the $j$-th Gaussian along the view ray:
\begin{equation}
\alpha_j(\mathbf{x}) 
= \sigma_j e^{-\frac{1}{2}(\mathbf{x} -\boldsymbol{\mu'}_j) \boldsymbol{\Sigma'} (\mathbf{x} - \boldsymbol{\mu'}_j)^T} .\end{equation}
%
Here $\boldsymbol{\mu'}$ and $\boldsymbol{\Sigma'}$ denote the projected 2D mean and covariance on the image plane, obtained by applying an affine approximation of the projective transform~\citep{zwicker2001EWA}.

\paragraph{3DGS Memory} Standard 3DGS pipelines store the full set of per-Gaussian attributes, training images, and optimizer state in GPU memory (VRAM); see Figure~\ref{fig:MemDistribution} for a detailed per-Gaussian breakdown. Additional temporary allocations occur during forward and backward passes (\eg, for sorting and gradient accumulation). This overhead varies with the scale of the Gaussians and effectiveness of culling strategies, but can be roughly upper-bounded by around 800 bytes per Gaussian in practice. This limits typical training to roughly $500\,000$ Gaussians per GB of GPU memory, imposing strict constraints on the detail and extent of reconstructions.

\paragraph{Gaussian Hierarchies} As introduced in H-3DGS \citep{Hierarchical}, Gaussian hierarchies recursively merge nearby Gaussians into a tree, where each non-leaf node approximates its children, and leaves correspond to the original Gaussians. A cut is defined by a condition $c_{\text{hier}}(i, \text{cam})$ evaluated in a breadth-first search (BFS). If a node satisfies this condition, it is added to the cut set and its children are skipped; otherwise, the BFS continues. A \emph{proper cut set} includes no parents or children of any included node and thus provides a view-adaptive LoD representation.

The cut condition used in \cite{Hierarchical} is a simple cut-off according camera distance:
\begin{equation}
c_{\text{hier}}(i, \text{cam}) = \left\|\boldsymbol{\mu}_i - \mathbf{p}_{\text{cam}}\right\|_2 \geq m_d(i), \quad m_d(i) = \frac{T}{\max_j s_i^j},
\end{equation}
where $\mathbf{p}_{cam}$ is the camera position, $T$ is a global LoD threshold, and $m_d(i)$ is the minimum acceptable distance for viewing Gaussian $i$. The BFS ensures that if $i$ is in the cut set, parent($i$) failed the condition: 
$
m_d(\text{parent}(i)) > \left\|\boldsymbol{\mu}_i - \mathbf{p}_{\text{cam}}\right\|_2 \geq m_d(i).
$


%

\section{Method}

To train models that exceed GPU memory limits, all Gaussian attributes are stored in CPU RAM and streamed to the GPU on demand for each training view according to the LoD hierarchy.
To accelerate the hierarchy cut, we store a copy of only the tree structure in VRAM, where larger subtrees are replaced by \textit{Sequential Point Trees} (SPTs), forming a \emph{hierarchical~SPT}~(HSPT).
To minimize costly transfers between RAM and VRAM, we track which SPTs currently reside in GPU memory and at which detail.
Only if an SPT is not present in this GPU cache at a sufficiently similar level of detail, will it be loaded from RAM.
Densification is performed on the CPU by adding new leaf nodes to the hierarchy and respawning low-opacity leaf nodes.
Following densification, the updated hierarchy is converted back to an HSPT and transferred to the GPU for a new round of training iterations.
\appref{App:Initialization} includes more details on initialization and training. 
An overview of our training and densification process can be found in Figure \ref{fig:Overview} and pseudocode in \appref{App:Pseudocode}.

\subsection{Sequential Point Trees for Gaussian Splatting}


\begin{figure}
    \centering
    \begin{subfigure}[b]{0.55\linewidth}
        \centering
        \includegraphics[width=1\linewidth]{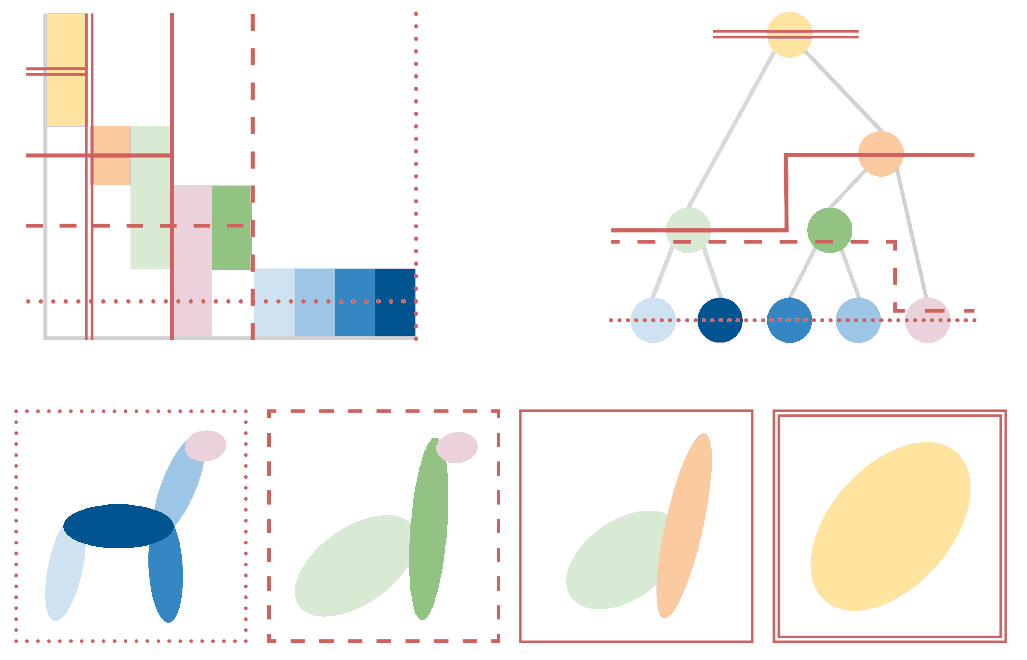}
        \caption{Hierarchy cuts} 
        \label{fig:Hierarchy_SPT}
    \end{subfigure}
    \hfill
    \begin{subfigure}[b]{0.43\linewidth}
        \centering
        \includegraphics[width=1\linewidth, trim=0 0 0 40, clip]{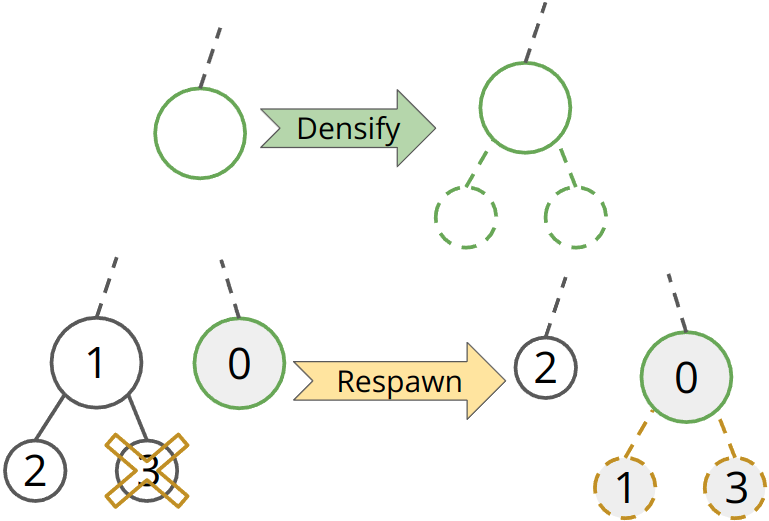}
        \caption{Densification example}
        \label{fig:Densification}
    \end{subfigure}
    
    \caption{\textbf{Visualization of Hierarchy and Densification.} 
    (\subref{fig:Hierarchy_SPT}) An SPT and Gaussian hierarchy show the same 5 Gaussians at varying levels of detail (red lines indicate cuts). Vertical lines show binary search results; horizontal lines show distance cuts. 
    (\subref{fig:Densification}) Leaf node densification and respawning.}
    \label{fig:h_spt}
\end{figure}%
Sequential Point Trees \citep{SPT} were originally developed for point cloud LoD rendering, but can be trivially extended to ellipsoids and Gaussians. They enforce a more constrained but more efficient cut condition than \citet{Hierarchical}:
\begin{equation}
\scriptsize
c_{\text{SPT}}(i, \text{cam}) = m_d(\text{parent}(i)) > \left\|\boldsymbol{\mu}_{\text{root}} - \mathbf{p}_{\text{cam}}\right\|_2 \geq m_d(i).
\end{equation}
This condition is evaluated for all Gaussians in parallel, using the shared root-camera distance $\left\|\boldsymbol{\mu}_{\text{root}} - \mathbf{p}_{\text{cam}}\right\|_2$. 
It requires storing only sorted pairs $\big(m_d(i), m_d(\text{parent}(i))\big)$, significantly reducing memory compared to full Gaussian hierarchies.
To optimize cuts, Gaussians are sorted by $m_d(\text{parent}(i))$ in descending order. A binary search determines the cutoff index $N$, above which Gaussians are too fine to be rendered.
Note that cuts are guaranteed to be proper, with nodes where $ m_d(\text{parent}(i))>m_d(i)$ never being selected for the cut.
Evidently, the level of detail of all Gaussians in the SPT is dictated by the camera's distance to its root node.
This can lead to Gaussians with $m_d(i) > \left\| \boldsymbol{\mu}_{i} - \mathbf{p}_{cam} \right\|_2$ being rendered, even though they would be too coarse for the current view.
To counteract this issue, we define:
$
M_d(i) = m_d(i) + \left\|\boldsymbol{\mu}_i - \mathbf{p}_{\text{cam}}\right\|_2,
$
as a conservative minimum distance function. By the triangle inequality, selecting Gaussians satisfying $M_d(i) \leq \left\|\boldsymbol{\mu}_{\text{root}} - \mathbf{p}_{\text{cam}}\right\|_2$ guarantees $m_d(i) \leq \left\|\boldsymbol{\mu}_i - \mathbf{p}_{\text{cam}}\right\|_2$.
In turn, this means that Gaussians that are further away from the camera than the root node will be selected at a higher level of detail.
SPTs are best suited for tightly grouped Gaussians observed from distances greater than their mutual spacing. Their compact memory footprint and parallel evaluation make them well-suited for large-scale scenes. Figure~\ref{fig:h_spt} visualizes both hierarchy types and their LoD cuts. 

\subsection{Densification}

Densifying an LoD representation presents a unique challenge, as the hierarchical structure must evolve continuously during training. Prior works circumvent this issue by constructing LoD hierarchies only after chunk-level training and densification are complete.

We take inspiration from \textit{3DGS-MCMC}~\citep{3DGSasMCMC}, which `splits' Gaussians, replacing them with two new Gaussians which together should appear similarly to the original.
Notably, this mirrors how a parent node in a Gaussian hierarchy approximates its children.
Motivated by this correspondence, we adopt this approach and instead `spawn' two new child nodes for a Gaussian, increasing the size of the hierarchy with minimal artifacts. This procedure leads to gradual increase in detail during densification, thereby avoiding the instability on large scenes observed in \citep{Hierarchical}. 

Instead of pruning, \textit{3DGS-MCMC} declares Gaussians below a certain opacity threshold as `dead', and respawns them at the position of a high-opacity Gaussian.
We propose a similar strategy: when a leaf node dies, its parent is replaced by its sibling node; the dead leaf node and its parent are then respawned as children to another node, which is selected to be densified.
See Figure \ref{fig:Densification} for an overview of the two hierarchy densification operations. Together, they ensure that the hierarchy can be expanded during training in a stable and valid manner while being rebalanced as required. 

While \textit{3DGS-MCMC} choose Gaussians to densify using a random selection weighted by opacity, we employ the strategy from \citet{Hierarchical}, which selects Gaussians for densification based on their maximal screen-space gradient.
This criterion better aligns densification with view-dependent reconstruction error in large-scale scenes.
For further details on densification, see \appref{App:Densification}.



\subsection{The Hierarchical SPT Datastructure}
We first review  previous LoD selection approaches and motivate the need for a new datastructure---the hierarchical SPT---for robust and efficient training. 
\paragraph{BFS} Computing the cut set of a large Gaussian hierarchy is costly and must be done for every frame. A straightforward solution is a BFS from the root, which guarantees a proper cut and enables early pruning of large subtrees (\eg, via frustum culling). However, graph traversal poorly suited to parallel execution on the GPU, making this approach prohibitively expensive at scale.

\paragraph{Parallel Cut} 
To enable GPU-accelerated cuts, \citet{Hierarchical} evaluate the cut condition in parallel for each Gaussian:
%
\begin{equation}
\scalebox{0.85}{%
    $\displaystyle \Big(m_d(i) < \left\| \boldsymbol{\mu}_{i} - \mathbf{p}_{cam}\right\|_2\Big)\land \Big(m_d(\text{parent}(i) )\geq \left\| \boldsymbol{\mu}_{\text{parent}(i)} - \mathbf{p}_{cam}\right\|_2 \Big)$,
}
\end{equation}
%
where any Gaussian that is sufficiently small at its current camera distance and whose parent is too large to be rendered, should be part of the cut set.
This produces a proper cut under the assumption that child Gaussians always have a smaller minimal distance than their parents (\ie the heap condition is fulfilled):
    $\forall i : m_d(i) < m_d(parent(i)) .$
This is generally valid when hierarchies are constructed after training, since parent Gaussians represent coarser approximations of their children. However, when the hierarchy is modified during training and densification, optimization can break the heap condition---leading to invalid cut sets and degenerate hierarchies that worsen over time.


\paragraph{Hierarchical SPT (HSPT)} 
\begin{figure}
    \centering
    \includegraphics[width=1.0\linewidth]{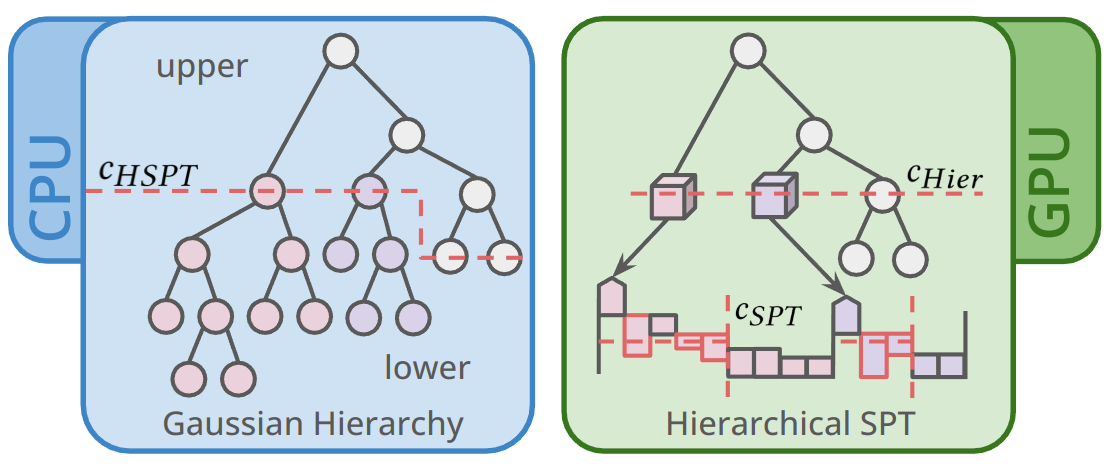}
    \caption{ A Gaussian hierarchy is converted to an HSPT by cutting according to Gaussian volume and converting sufficiently large subtrees to SPTs. The HSPT can then be cut in a 2-step process.}
    \label{fig:HierarchicalSPT}
    \end{figure}      
Our HSPT data structure combines the benefits of both approaches. 
To construct it from a Gaussian hierarchy, we cut it using a BFS on the condition
$c_\text{HSPT}(i) = s_i^1\cdot s_i^2 \cdot s_i^3 < \texttt{size}$
with volume threshold $\texttt{size}$. The resulting cut set $\mathbb{C}_{\text{HSPT}}$ partitions the hierarchy into the \emph{upper hierarchy}, which includes all Gaussians with volume greater than $\texttt{size}$, and the \emph{lower hierarchy}, consisting of the subtrees rooted at the nodes in the cut set.

The volume of each root node in the lower hierarchy is now bounded by $\texttt{size}$, which also roughly bounds the extent of all Gaussians in the subtree. This provides an upper bound on the error introduced if the subtree is converted into an SPT. Consequently, each subtree of sufficient size in $\mathbb{C}_{\text{HSPT}}$ can be transformed into an SPT to accelerate cut computation without violating cut correctness at higher levels.
The HSPT-based cutting process then proceeds in two steps: first, a BFS on the upper hierarchy selects the required nodes and leaf/SPT subtrees for the current view. Second, each selected SPT is cut according to the camera's distance to its root node. Together, these yield the full set of Gaussians needed for rendering the current frame.
The construction and cutting process of an HSPT is illustrated in Figure~\ref{fig:HierarchicalSPT}. Figure~\ref{fig:ContinuousLOD} and \ref{fig:SPTs} show example frames with respectively SPTs highlighted and different levels of detail. 

Rebuilding the HSPT every training iteration would eliminate its performance benefits. Instead, we exploit the fact that the minimum distance $m_d$ evolves slowly during optimization and thus requires only infrequent updates. In practice, we rebuild the HSPT only after each densification step.
This infrequent recomputation allows us to use a more accurate---albeit more expensive---minimum distance metric than the inverse of maximal scale. Specifically, we define:
\begin{equation}
    m_d'(i) = \frac{T}{\sqrt{s_i^1 \cdot s_i^2 + s_i^1 \cdot s_i^3 + s_i^2 \cdot s_i^3}},
\end{equation}
which corresponds to the inverse square root of the surface area of the Gaussian ellipsoid (up to a constant factor). This better captures the perceived size of anisotropic Gaussians, especially those that are significantly elongated in one or more directions.

\paragraph{Frustum Culling}
The main benefits of using BFS to cut the upper hierarchy are the guarantee of a proper cut and early culling of subtrees. We therefore frustum cull every node considered in the BFS by checking if a sphere around the Gaussians with radius $3 \cdot \max_j s_i^j$ intersects the view frustum.
While using the Gaussian scale as a conservative proxy for the entire subtree extent is not perfectly accurate, we observed no discernable difference compared to a full bounding sphere hierarchy in our experiments. 
Although Gaussians are implicitly frustum culled during rasterization, this early culling accelerates cut computation and significantly reduces the number of Gaussians loaded from RAM (see Figure \ref{fig:FrustumCull}). 

\subsection{Caching on the GPU}

Loading Gaussian data from RAM is a costly operation that can become a significant bottleneck during large-scale training. To mitigate this, we maintain a GPU-resident cache of Gaussians that are likely to be reused across consecutive training views. However, checking the cache for every individual Gaussian would introduce non-trivial overhead. Once again, SPTs offer an efficient alternative.

Rather than caching individual Gaussians, we store the Gaussians from SPT cuts along with the cached distance from the camera to the root of each SPT, denoted $\bar{d}^j$ for the $j$-th SPT. During rendering, when the upper hierarchy is cut and the required SPTs identified, we compute 
$d^j = \|\boldsymbol{\mu}_{\text{root}(j)} - \mathbf{p}_{\text{cam}}\|_2$ and check whether a matching cut is cached, using a simple distance ratio tolerance:
\begin{figure}
    \includegraphics[width=0.8\linewidth]{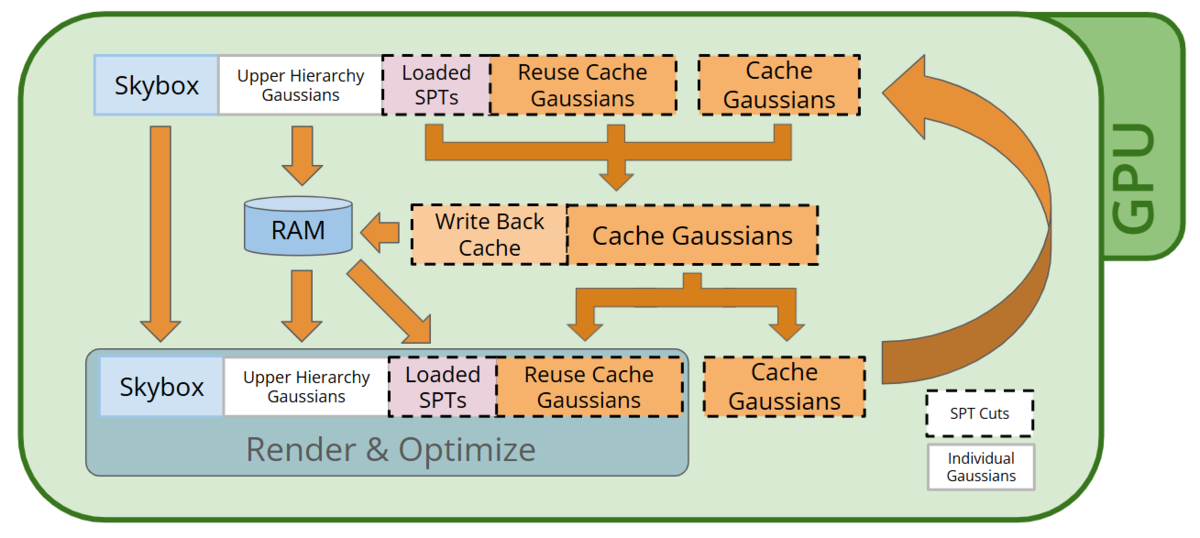}
    \caption{Gaussians required for the current training view are assembled from three sources: the upper tree, newly loaded SPT cuts from RAM, and cache hits. After optimization, newly accessed SPTs are added to the GPU cache.}
    \label{fig:Caching}
\end{figure}
\begin{equation}
    D_{\text{min}} \leq \frac{d^j}{\bar{d}^j} \leq D_{\text{max}}.
\end{equation}
Here, $D_{\text{max}}$ defines the allowable range for using coarser-than-optimal detail, while $D_{\text{min}}$ limits how much finer detail can be tolerated. If this condition is met, the cached SPT cut is reused, avoiding a costly RAM-to-GPU transfer.

While this heuristic introduces slight variability in rendered detail---since the LoD may depend on the cache state---we find that this stochasticity actually improves training robustness. In particular, subtle variations in detail across views discourage overfitting to fixed camera distances and promotes generalization across scales.

For each training view, visible Gaussians are assembled from the upper hierarchy, the cached SPTs, and the skybox Gaussians (which remains in VRAM). Uncached SPTs are streamed from RAM. After each training iteration, newly loaded SPTs are added to the cache.

To bound VRAM usage, we use a least-recently-used (LRU) write-back policy. When a memory threshold is exceeded, entries are written back to RAM. Additionally, to prevent overfitting to persistent cache entries, the entire cache is flushed every $1\,000$ iterations. Figure~\ref{fig:Caching} illustrates the caching process across two frames.

\paragraph{View Selection}
In large-scale scenes, the GPU cache typically covers only a small fraction of the overall geometry, leading to sparse cache hits. To improve cache utilization, we prioritize spatial locality by selecting successive training views close to the current one, maximizing Gaussian reuse.

To this end, we precompute a $k$-nearest-neighbour graph over all training view positions, where edge weights $w_{ij}$ correspond to the Euclidean distance between views $i$ and $j$. The next training view $j$ is then sampled from the $k$-nearest neighbours of the current view $i$ according to the distribution:
$
\mathbb{P}(j \mid i) \propto \frac{1}{w_{ij} + W},
$
where $W$ is a normalization constant that also controls the degree of exploration.
However, care must be taken when deviating from a uniformly sampled training view selection as this may introduce bias. To counteract this, we inject a randomly selected view every 128 iterations, which we find sufficient to preserve generalization performance.


\subsection{Memory Layout}

\begin{figure}
    \centering
    \includegraphics[width=1.0\linewidth]{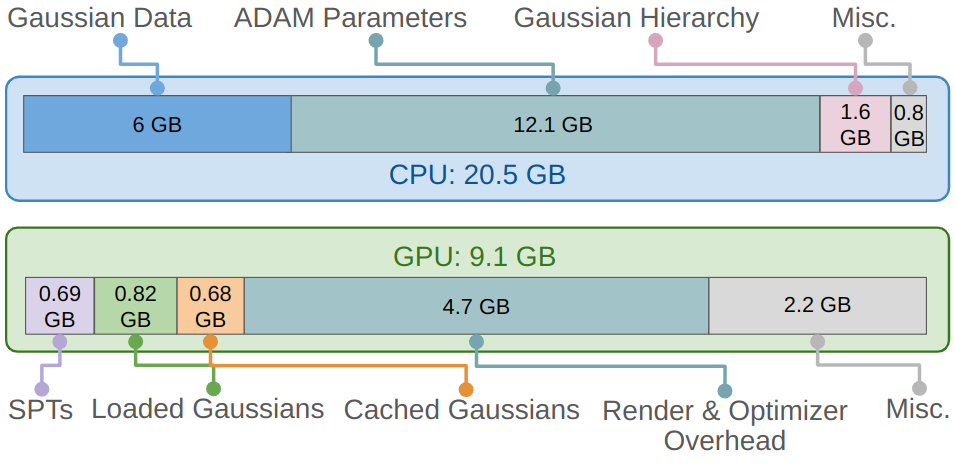}
    
    \caption{Peak memory consumption of CPU and GPU for a training iteration on MC-smaller-city+ with 60 million Gaussians.}
    \label{fig:MemLayout}
\end{figure}
Figure~\ref{fig:MemLayout} illustrates the peak memory usage for a single training iteration of a 60-million-Gaussian hierarchy on the MC-smaller-city+ dataset. 
The majority of RAM usage is consumed by per-Gaussian properties and their corresponding ADAM optimizer states. In contrast, the hierarchy structure itself accounts for less than 10\% of the total RAM footprint. On the GPU, the SPT metadata for all 60 million Gaussians occupies just 680\,MB of VRAM and the upper hierarchy negligible 24\,MB.

Even in wide-angle aerial views, only a subset of the scene is actively loaded into GPU memory. In the example shown, 2.2 million Gaussians are rendered directly, while an additional 2.4 million are retained in the cache for future use. The bulk of GPU memory is instead consumed by temporary allocations for rasterization and optimization, which scale with the number of Gaussians rendered. Therefore, minimizing the number of active Gaussians is critical for staying within the VRAM budget.

The remaining GPU memory usage consists of auxiliary data, including cache management, hierarchy cut tracking, training and ground-truth images, as well as general PyTorch overhead. Figure~\ref{fig:MemDistribution} provides a breakdown of all data associated with a single Gaussian and how they are distributed across CPU and GPU.



\begin{table}[t]
    \centering
        \caption{
\textbf{Novel view synthesis results}. Results with \textsuperscript{\textdagger} require suboptimal COLMAP initialization instead of the provided point cloud. VRAM usage is measured while rendering the test images. Methods are considered out-of-memory (OOM) if they exceed 141GB of VRAM during training or rendering. 
 }
\setlength{\tabcolsep}{2pt}%
\resizebox{1\linewidth}{!}{
\begin{tabular}{lrrrrrrr}
\toprule
 
Method & PSNR\textsuperscript{$\uparrow$} & SSIM\textsuperscript{$\uparrow$} & LPIPS\textsuperscript{$\downarrow$} & VRAM (render)\textsuperscript{$\downarrow$}& VRAM (train)\textsuperscript{$\downarrow$} & \#iterations\textsuperscript{$\downarrow$} & \#Gaussians \\
\midrule
\multicolumn{8}{c}{MC-small-city+ (42.2k images) } \\
\cmidrule{1-8}
CityGaussian & \cellcolor{blue!10}19.78& \cellcolor{blue!10}0.650 & \cellcolor{blue!10}0.475   &\cellcolor{blue!25} 2.4GB & \cellcolor{blue!25} 4.1GB & 1.11M & 2.7M\\
HorizonGS & 12.06 & 0.521 & 0.544   & 20.6GB & 25.3GB & 2.5M & N/A \\ 
OctreeGS & OOM & OOM & OOM & OOM   & OOM  & OOM & OOM \\
H-3DGS\textsuperscript{\textdagger} & OOM & OOM  & OOM  & OOM & OOM & OOM & OOM \\
CLM & 15.68 & 0.584 & 0.523 &  18.7GB  & 25.1GB & \cellcolor{blue!25} 600k & 24.2M \\
Ours & \cellcolor{blue!25}21.59 & \cellcolor{blue!25} 0.711  &\cellcolor{blue!25} 0.396  & \cellcolor{blue!10} 16.4GB  & \cellcolor{blue!10} 18.0GB & \cellcolor{blue!25} 600k & 136.7M \\
\midrule
 \multicolumn{8}{c}{MC-smaller-city+ (15.1k images)} \\
\cmidrule{1-8}
HorizonGS & 15.28 & 0.601 & 0.435  & \cellcolor{blue!25} 13.9GB 
 & 
 23.6GB & 1.2M & 34.1M \\
OctreeGS & 15.93 & 0.560 & 0.556  & 26.6GB & 98.7GB & \cellcolor{blue!25} 310k & N/A\\
Ours & \cellcolor{blue!25} 21.71 & \cellcolor{blue!25} 0.710 & \cellcolor{blue!25} 0.370   &  15.7GB & \cellcolor{blue!25}17.2GB & \cellcolor{blue!25} 310k & 60M  \\
\midrule
H-3DGS\textsuperscript{\textdagger} & 14.57 & 0.516  & 0.563  & 97.0GB & 15.1GB \cellcolor{blue!25}& 4M & 83.1M \\
Ours\textsuperscript{\textdagger}& \cellcolor{blue!25} 20.59 & \cellcolor{blue!25} 0.658 & \cellcolor{blue!25} 0.468 & \cellcolor{blue!25} 20.7GB & 22.1GB & \cellcolor{blue!25} 310k  & 24.7M  \\
\midrule
 \multicolumn{8}{c}{Uni10k (10.1k images)} \\
\cmidrule{1-8}

CityGaussian  & 19.78 & 0.642 & 0.414  &\cellcolor{blue!25} 3.7GB & \cellcolor{blue!25} 14.6GB & 300k & 2.2M\\
H-3DGS  &   19.96 & 0.705 & 0.319 & 39.2GB  & 26.5GB & 255k & 32.5M \\

OctreeGS &  20.06 & 0.602 & 0.341 & 25.3GB &\cellcolor{blue!10} 22.9GB & \cellcolor{blue!25} 160k & N/A\\
CLM-GS & 19.89 & 0.608 & 0.455 & 46.7GB & 39.7GB & \cellcolor{blue!25} 160k & 12.6M\\
HorizonGS  &   \cellcolor{blue!25}21.72 &  \cellcolor{blue!25}0.722 &  \cellcolor{blue!25}0.326  & \cellcolor{blue!10}10.3GB & 44.3GB & 400k & N/A 
\\
Ours &   \cellcolor{blue!10}21.32 & \cellcolor{blue!10} 0.718 &  \cellcolor{blue!10} 0.336  & 22.6GB & 31.3GB& \cellcolor{blue!25} 160k & 28.6M\\
\bottomrule
\end{tabular}
}
\label{tab:Results}
\end{table}


\section{Evaluation}
Our method is designed to enable seamless training and rendering on ultra-large-scale scenes comprising tens of thousands of views captured at vastly different scales. Unfortunately, most datasets of sufficient size contain either street-level or aerial views---but not both. To address this gap, we captured the campus of Udine University with over 10\,000 images at 4k resolution from varying aerial heights and street views in the Uni10k scene.
We also introduce MC-small-city+, which expands the small-city scene from the MatrixCity~\citep{MatrixCityDataset} dataset with new high-altitude aerial views covering the entire scene, resulting in 42.2k images spanning hundreds of buildings.
Since some methods run out-of-memory on scenes of this scale, we additionally construct a subset of MC-small-city+ covering about a third of the area (15.1k images, MC-smaller-city+).
Additionally, we present results on large-scale street-view and indoor scenes from the Hierarchical 3DGS~\citep{Hierarchical} and OccluGaussian~\citep{OccluGaussian} datasets. Results on the aerial datasets UrbanScene3D~\citep{UrbanScene3D} and Mill19~\citep{MegaNerf}, as well as the smaller-scale H-3DGS single chunk scene are included in \appref{App:Datasets} along with additional details on all scenes.    

We choose recent divide-and-conquer based 3DGS methods \emph{CityGaussian} \citep{CityGaussian}, \emph{H-3DGS} \citep{Hierarchical}, the large-scale neural Gaussian method \emph{OctreeGS} \citep{OctreeGS} and out-of-core training method \emph{CLM-GS} \citep{CLM} as baselines. Because \emph{OccluGaussian} \citep{OccluGaussian} has not released code yet, we compare only against their self-reported results on their dataset.
To enable training on scenes of this scale, it was necessary to modify \emph{CityGaussian}, \emph{HorizonGS} and \emph{OctreeGS} to load images from disk instead of caching them fully in RAM or VRAM. 
These changes affect training throughput but do not alter optimization behavior or final reconstruction quality.
In general, the hyperparameters suggested for large-scale scenes were used for the experiments. Exact details on the training setting and configuration files are included in the supplemental material.
Evaluations were run on a single H200 GPU with 141 GB VRAM to allow baselines methods to complete training.
Where applicable, we report rendering performance on consumer GPUs to demonstrate practical deployability.

\subsection{Results}
Qualitative comparisons are shown in Figure \ref{fig:Quality} and quantitative comparisons in Tables~\ref{tab:Results} and \ref{tab:H3DGS}. 
The outputs of \emph{H-3DGS} show significant floating and ghosting artifacts. While \emph{H-3DGS} performs well on individual chunks of MC-smaller-city+, oversized floaters survive the merging procedure and obscure the majority of test images  (cf. Figure~\ref{fig:artifacts}). Floaters and ghosting artifacts can also be found in the results on other scenes. The merging procedure of \emph{CityGaussian}, which is designed for aerial-only datasets,  discards most of the trained Gaussians to avoid chunk artifacts, resulting in mostly artifact-free views and low memory requirements, but noticeably blurry results.  
\emph{A LoD of Gaussians} reconstructs novel views for joined street- and aerial data with a high degree of detail while avoiding chunk-based artifacts, resulting in a consistent improvement in quality metrics across scenes. 
Seamless training also enables faster convergence, substantially reducing the number of training iterations required compared to  divide-and-conquer based methods. Results on the OccluGaussian dataset further demonstrate generalizability to indoor environments, which are particularly difficult for divide-and-conquer based methods. 
All experiments of our method were run on an RTX 3090 GPU, except Uni10k, which required more VRAM for the 4k training images. Lowering the resolution to HD reduces peak VRAM to 20GB for the same parameters.

\emph{CLM-GS} also employs out-of-core memory during training, but relies solely on frustum culling to limit memory pressure. On Uni10k and MC-small-city+, this results in substantially higher VRAM usage per Gaussian than our method.
Further, unlike \emph{OctreeGS}~\citep{OctreeGS} and our approach, their densification is not tailored to large-scale scene reconstruction from sparse views and point clouds. On MatrixCity, \emph{CLM-GS} instead initializes from a fully dense synthetic point cloud; larger non-aerial scenes (\eg Campus) fail to reconstruct under all configurations provided with their code.

\emph{HorizonGS} \citep{HorizonGS} is not competitive on the MC-city+ tests due to the large scale of the dataset and its diversity of viewpoints. 
While the authors report results for MatrixCity, only a single block is used in their evaluation.
However, it performs competitively on Uni10k, for which it is explicitly designed: the method targets mixed street- and aerial-view scenarios and relies on ground-truth labels to primarily target street-level views for densification.
Our universal approach---without additional supervision---closely matches \emph{HorizonGS} on Uni10k and drastically outperforms it on both MC-city+ scenes. 
Furthermore, rendering the neural \emph{HorizonGS} representation is impractical due to a costly preprocessing step: while enabling low VRAM usage, it results in render times of up to 10 seconds per frame.

\begin{table}[ht!]
\centering
\caption{
\textbf{H-3DGS and OccluGaussian dataset novel view synthesis results.}
Results with \textsuperscript{\textdagger} are taken from \cite{OccluGaussian}, parentheses contain the number of images per-scene.
We strongly suspect they evaluated LPIPS using the AlexNet backbone, so we report the same for our results as LPIPS$_\text{A}$.
}
\setlength{\tabcolsep}{2pt}%
\resizebox{1\linewidth}{!}{
\begin{tabular}{lrrrrrrrrrrrr}
\toprule
Dataset & \multicolumn{6}{c}{Hierarchical 3DGS} & \multicolumn{6}{c}{OccluGaussian} \\
\cmidrule(lr){2-7}\cmidrule(lr){8-13}
Scene & \multicolumn{3}{c}{Small City (5.8k)} & \multicolumn{3}{c}{Campus (22.0k)} & \multicolumn{3}{c}{Canteen (8.6k)} & \multicolumn{3}{c}{Classbuilding (9.1k)} \\
\cmidrule(lr){2-4}\cmidrule(lr){5-7}\cmidrule(lr){8-10}\cmidrule(lr){11-13}
Method & PSNR\textsuperscript{$\uparrow$} & SSIM\textsuperscript{$\uparrow$} & LPIPS\textsuperscript{$\downarrow$} & PSNR\textsuperscript{$\uparrow$} & SSIM\textsuperscript{$\uparrow$} & LPIPS\textsuperscript{$\downarrow$} & PSNR\textsuperscript{$\uparrow$} & SSIM\textsuperscript{$\uparrow$} & LPIPS$_\text{A}$\textsuperscript{$\downarrow$} & PSNR\textsuperscript{$\uparrow$} & SSIM\textsuperscript{$\uparrow$} & LPIPS$_\text{A}$\textsuperscript{$\downarrow$} \\
\midrule
OctreeGS & 19.20 & 0.601 & 0.498 & \cellcolor{blue!10} 20.54 &  0.643 &  0.544  & 21.49\phantom{\textsuperscript{\textdagger}} & 0.826\phantom{\textsuperscript{\textdagger}} & 0.293\phantom{\textsuperscript{\textdagger}} & 23.08\phantom{\textsuperscript{\textdagger}} & 0.889\phantom{\textsuperscript{\textdagger}} & 0.197\phantom{\textsuperscript{\textdagger}} \\
CityGS & 21.40 & 0.692 & 0.383   & 19.47 & \cellcolor{blue!10}0.645 & 0.531 & 20.41\textsuperscript{\textdagger} & 0.794\textsuperscript{\textdagger} & 0.275\textsuperscript{\textdagger} & 20.48\textsuperscript{\textdagger} & 0.840\textsuperscript{\textdagger} & 0.244\textsuperscript{\textdagger} \\
H-3DGS  & \cellcolor{blue!10} 24.35 & \cellcolor{blue!25} 0.788 & \cellcolor{blue!25} 0.273  
& 17.84 & 0.610 & \cellcolor{blue!10} 0.467 & 22.71\textsuperscript{\textdagger} & 0.825\textsuperscript{\textdagger} & 0.178\textsuperscript{\textdagger} & 23.87\textsuperscript{\textdagger} & 0.881\textsuperscript{\textdagger} & 0.128\textsuperscript{\textdagger}  \\ 
CLM-GS & 22.54 & 0.699 &  0.392  & 16.60 & 0.613 & 0.558 & 23.13\phantom{\textsuperscript{\textdagger}} & 0.864\phantom{\textsuperscript{\textdagger}} & 0.168\phantom{\textsuperscript{\textdagger}} & 22.89\phantom{\textsuperscript{\textdagger}} & 0.892\phantom{\textsuperscript{\textdagger}} & 0.159\phantom{\textsuperscript{\textdagger}} \\ 
OccluGS & N/A & N/A & N/A & N/A & N/A & N/A & \cellcolor{blue!25} 25.25\textsuperscript{\textdagger} & \cellcolor{blue!25}0.900\textsuperscript{\textdagger} & \cellcolor{blue!25}0.100\textsuperscript{\textdagger} & \cellcolor{blue!25}25.33\textsuperscript{\textdagger} & \cellcolor{blue!25}0.921\textsuperscript{\textdagger} & \cellcolor{blue!25}0.083\textsuperscript{\textdagger} \\
Ours  & \cellcolor{blue!25} 24.61 &\cellcolor{blue!10} 0.774 & \cellcolor{blue!10}  0.297 & \cellcolor{blue!25} 21.85 & \cellcolor{blue!25} 0.707 & \cellcolor{blue!25} 0.456 & \cellcolor{blue!10}23.33\phantom{\textsuperscript{\textdagger}} & \cellcolor{blue!10}0.867\phantom{\textsuperscript{\textdagger}} & \cellcolor{blue!10}  0.155\phantom{\textsuperscript{\textdagger}} & \cellcolor{blue!10}24.14\phantom{\textsuperscript{\textdagger}} & \cellcolor{blue!10}0.915\phantom{\textsuperscript{\textdagger}} &\cellcolor{blue!10}  0.116\phantom{\textsuperscript{\textdagger}} \\  
\bottomrule
\end{tabular}
}
\label{tab:H3DGS}
\end{table}

\paragraph{Rendering} Our level-of-detail and caching strategy can also be applied to efficiently render the trained models. As supplemental material, we include fly-through videos of the evaluated scenes rendered on an RTX 3090. Table \ref{tab:Results} demonstrates effective VRAM reduction during rendering of test images compared to other 3DGS methods. Figure \ref{fig:RenderComp} compares rendering VRAM consumption and image quality of our LoD method with full-detail 3DGS \citep{3DGS} and gsplat \citep{Gsplat}. Our approach achieves visual fidelity comparable to the baselines while significantly reducing VRAM usage, highlighting the effectiveness of our LoD scheme.

\paragraph{Ablations}
We assess the contribution of key components through an ablation study using recorded camera paths across a selection of scenes (see supplemental videos). Table~\ref{tab:Ablations} reports average frame times over these paths.
Caching significantly improves rendering performance, roughly doubling the framerate across scenes by reducing the average number of Gaussians loaded from RAM by 93\% on Campus and 86\% on MC-small-city+. The effectiveness of frustum culling scales with scene size: On the full MC-small-city+ scene, 24.5 million Gaussians are frustum culled on average (88\% reduction), while for Campus and Small City the corresponding values are 9.8 million and and 7.9 million (74\% and 65\% reduction), respectively.
As expected, the overhead of frustum culling amortizes with scene size, making it essential for the largest datasets.

To evaluate cut efficiency, we compare the time required to compute the visible set using either full hierarchy BFS or our HSPT-based approach. HSPT consistently yields faster cut times due to improved parallelization. 
Moreover, the BFS approach requires positions and scales for all Gaussians to reside in memory, causing it to exceed 24GB of VRAM on MC-small-city+, whereas the HSPT method peaks at 21GB. 
For training ablations, we measure average iteration durations over 1\,000 steps. Here, frustum culling and caching prove essential, substantially reducing the number of Gaussians loaded and rendered per view. 

\begin{table}[h]
    \caption{\textbf{Ablations.} Average frame times for rendering camera paths and average iteration times during training with and without caching Gaussians. The final results show the average timings of the hierarchy cut during rendering using our HSPT and the baseline BFS approach. For Campus, we evaluate two different models with 38M and 80M Gaussians respectively.}
    \centering
\resizebox{0.9\linewidth}{!}{
    \centering
    \begin{tabular}{lcccc}
    \toprule
     & MC-smaller-city+ & MC-small-city+ & \multicolumn{2}{c}{Campus}\\
     & 60M & 136M & 38M & 80M  \\
    \midrule
       Render Full & \cellcolor{blue!25} 48.1 ms  & \cellcolor{blue!25}  68.3 ms 
        & 47.1 ms   
        & \cellcolor{blue!25} 83.2 ms
         \\ 
       Render w/o Cache &  119.4 ms& 127.4 ms  
       & 92.6 ms 
       & 222.3 ms \\ 
         Render w/o Frustum Culling & 52.3 ms & 89.3 ms & \cellcolor{blue!25} 38.3 ms & 110.2 ms \\

    \midrule
        Training Full & \cellcolor{blue!25} 156 ms & \cellcolor{blue!25} 174 ms  &  
 \multicolumn{2}{c}{\cellcolor{blue!25} 205 ms} \\
        Training w/o Cache & 471 ms & 257 ms  & \multicolumn{2}{c}{244 ms}  \\
        Training w/o Frustum Culling  & 685 ms
 & 626  ms & \multicolumn{2}{c}{312 ms} \\
                 \midrule
    Render Cut (HSPT) & \cellcolor{blue!25} 31.9 ms & \cellcolor{blue!25}  45.7ms   & \cellcolor{blue!25} 31.3 ms & \cellcolor{blue!25}  36.5 ms \\
        Render Cut (BFS) &   47.8 ms & OOM & 40.0 ms & 53.7 ms \\
    \bottomrule
    \end{tabular}
    }
    \label{tab:Ablations}
\end{table}
\label{App:Ablation}

\section{Discussion and Outlook}
\emph{A LoD of Gaussians} enables seamless training and rendering of ultra-large 3DGS models on consumer hardware. By storing Gaussian data in external memory and streaming it on demand, our method avoids the pitfalls of chunk-based pipelines. The HSPT datastructure accelerates LoD selection and remains robust to ongoing training changes. Combined with caching and view selection, our approach significantly reduces out-of-core overhead. These components enable efficient reconstruction and rendering at scale, as demonstrated on challenging multi-scale scenes such as MC-small-city+.

\paragraph{Limitations and Future Work}
Our method represents an informed trade-off between performance and memory. While it greatly reduces the necessary number of training iterations for large-scale scenes, individual iterations take longer than standard 3DGS training due to data loading and hierarchy management overhead.

Similarly, although rendering framerates are a significant improvement over neural-based and other out-of-core methods, they do not yet reach the peak performance of fully in-core 3DGS.

Our approach requires roughly 1~GB of RAM per million Gaussians, which---while more efficient than prior methods---still constrains scalability for extremely large scenes. Loading from disk is feasible in our experiments, but at the cost of about a $10\times$ slowdown, making fast secondary storage highly desirable.

The level-of-detail system makes our method robust to large variations in view distance.
However, when such variation is absent---for example in single-height aerial datasets as evaluated in \appref{App:Datasets}---the LoD machinery introduces unnecessary overhead, making more straightforward training a competitive alternative.
Interactive rendering performance could be further improved by avoiding per-frame hierarchy cut recomputation and asynchronous streaming. 

While frustum culling effectively reduces memory load in most views, it is ineffective when the entire scene lies inside the frustum. Occlusion culling could address this limitation by skipping entire SPTs before they are loaded into memory.

Overall, we believe that out-of-core 3D Gaussian Splatting is a promising direction for scaling radiance field methods to city-scale scenes and beyond on consumer-grade hardware.

\clearpage
\bibliographystyle{ACM-Reference-Format}
\bibliography{library}
\clearpage

\begin{figure*}[h]
\centering
\begin{minipage}{0.55\textwidth}
\begin{subfigure}{1\linewidth}
\centering
    \includegraphics[width=1\linewidth]{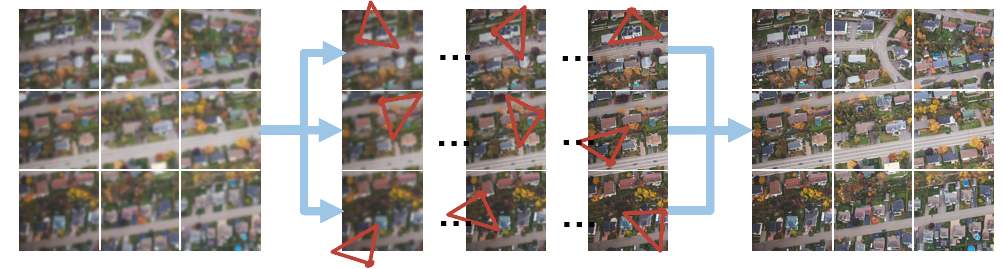}
    \caption{Divide and Conquer}
\end{subfigure}
\begin{subfigure}{1\linewidth}
\centering
    \includegraphics[width=1\linewidth]{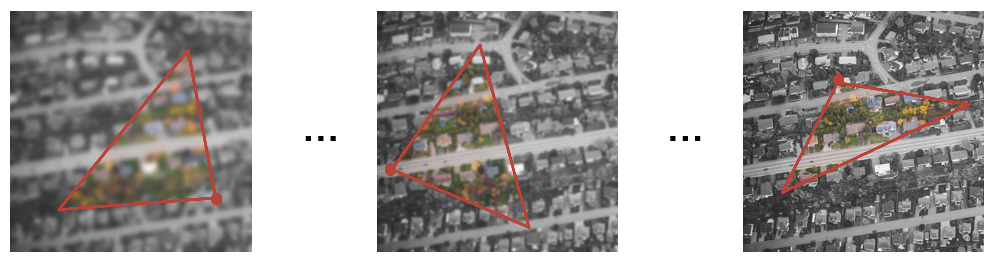}
    \caption{Ours}
\end{subfigure}
\caption{Comparison between the divide and conquer training process as used in \citet{Hierarchical, VastGaussian, CityGaussian} to our training process. Colored regions are present in VRAM, training views are drawn in red.}
\label{fig:Training}
\end{minipage}
\hspace{1cm}
\begin{minipage}{0.3\textwidth}
    \centering
    \begin{subfigure}{0.9\linewidth}
    \centering
        \includegraphics[width=0.7\linewidth]{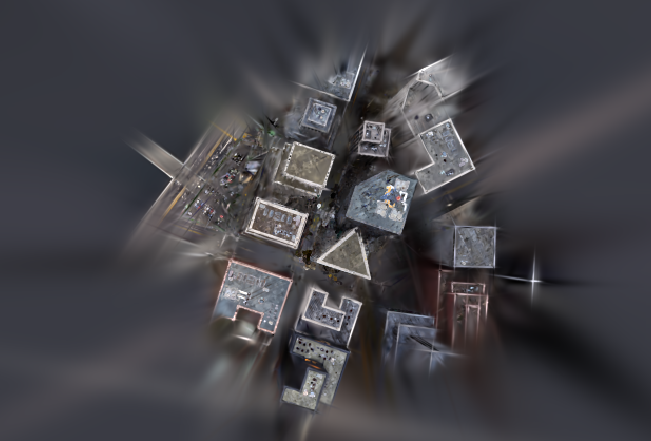}    
        \caption{Chunk Bleeding in \emph{H-3DGS} \citep{Hierarchical}.}
    \end{subfigure}
    \begin{subfigure}{0.9\linewidth}
    \centering
        \includegraphics[width=0.7\linewidth,trim=9.5cm 0 0 0,clip]{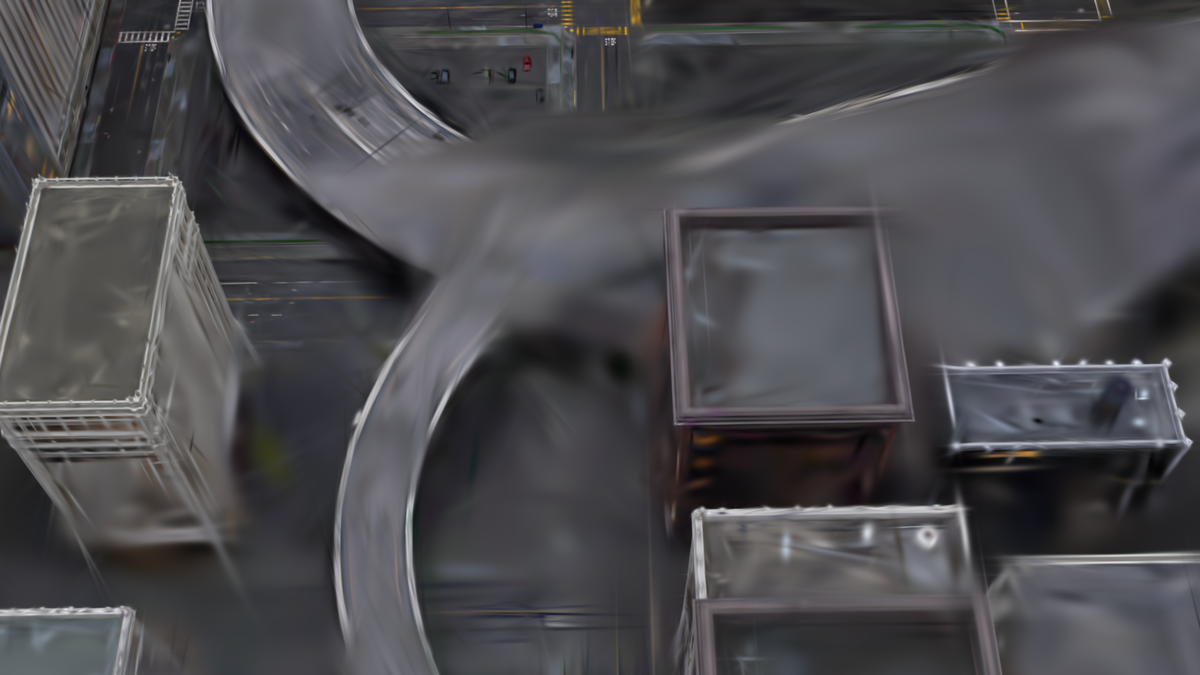}
        \caption{Chunk Ghosting in \emph{CityGaussian} \citep{CityGaussian}.}
    \end{subfigure}
    \caption{ Artifacts caused by the divide-and-conquer strategy on MC-small-city+.}
    \label{fig:artifacts}
    \end{minipage}
\end{figure*}

\begin{figure*}[h]
\begin{minipage}{0.63\textwidth}
    \centering
    \vspace{-0.3cm}
    \includegraphics[width=1\linewidth]{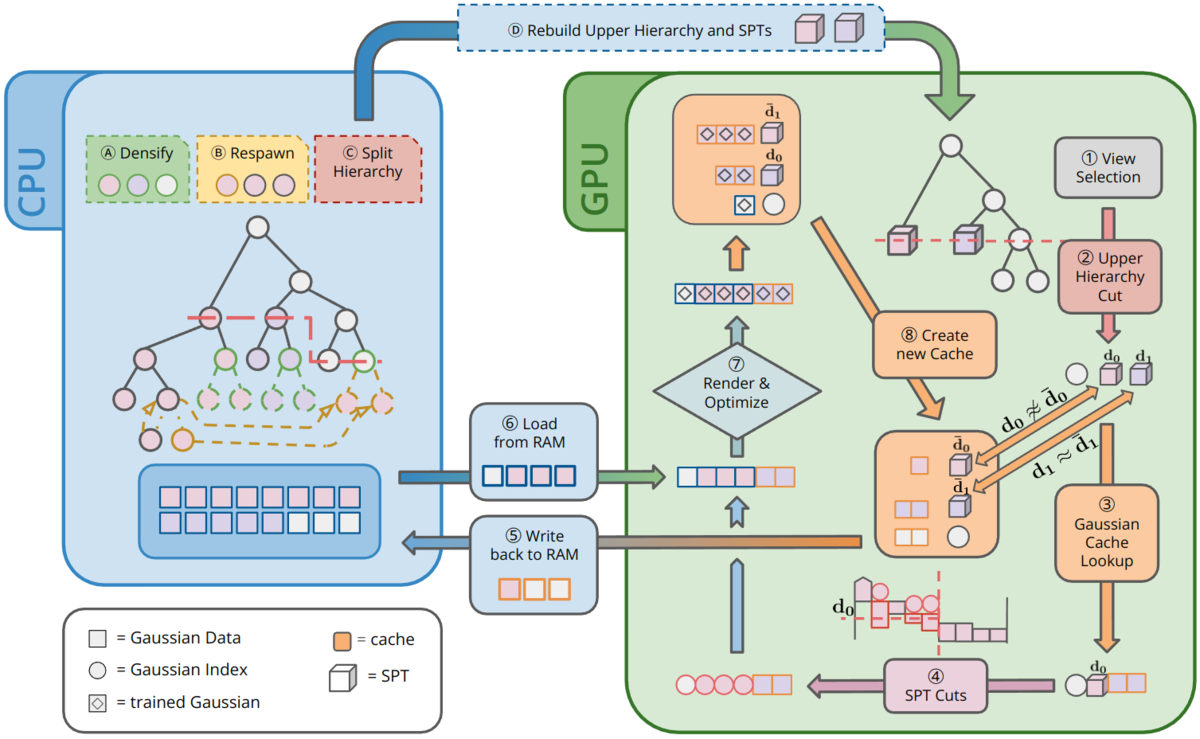}
    \caption{Method Overview: Steps \raisebox{.5pt}{\textcircled{\raisebox{-.9pt} {1}}} to \raisebox{.5pt}{\textcircled{\raisebox{-.9pt} {8}}} show the process of a single training iteration, while \raisebox{.5pt}{\textcircled{\raisebox{-.9pt} {A}}} through \raisebox{.5pt}{\textcircled{\raisebox{-.9pt} {D}}} show a densification step.}
    \label{fig:Overview}
\end{minipage}
\hfill
    \centering
    \vspace{-0.3cm}
    \begin{minipage}{0.33\linewidth}
        \centering

    \includegraphics[width=1\linewidth]{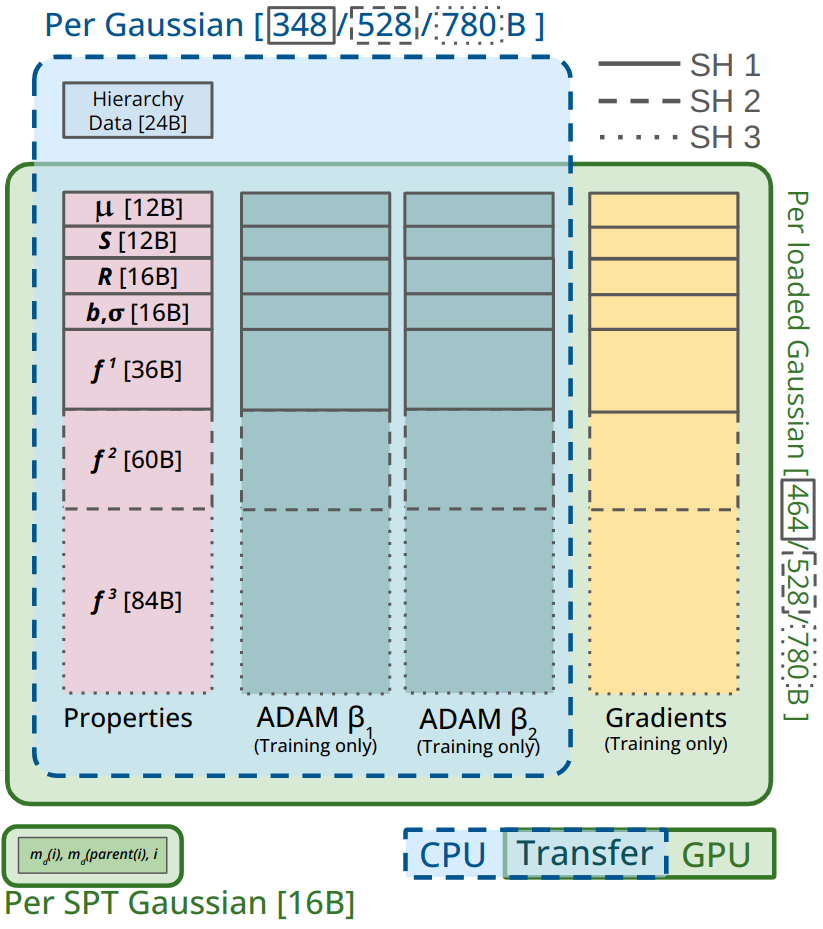}
    \caption{
    Memory Layout: Only the currently loaded Gaussians and slim SPT information needs to be stored on GPU.
    }
    \label{fig:MemDistribution}
    \end{minipage}
    \hspace{0.5cm}
    
\end{figure*}

\begin{figure*}[h]
\centering
\hfill
\begin{minipage}{.45\linewidth}
\centering
    \hfill\includegraphics[width=1\linewidth]{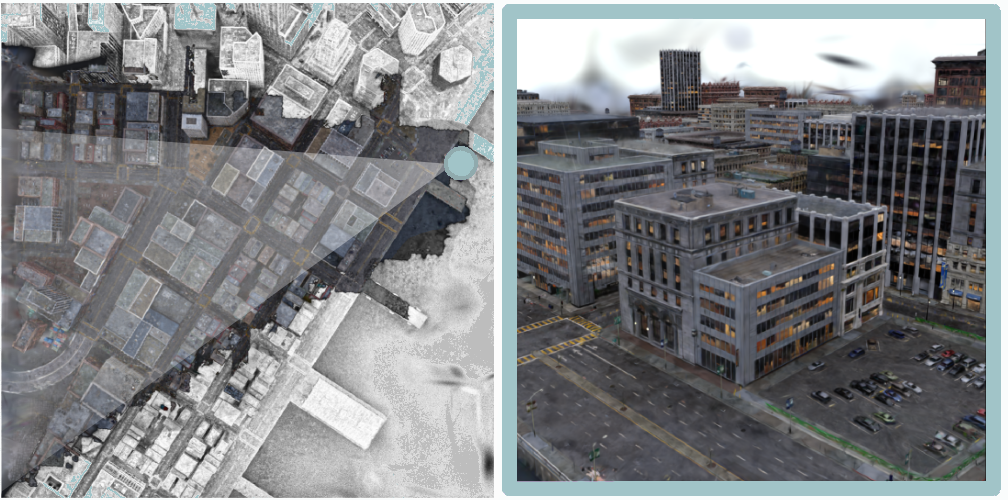}
    \captionof{figure}{Frustum Culling and LoD selection (left) greatly reduces the number of Gaussians required to render a view (right). \label{fig:FrustumCull}}
\end{minipage}\hfill
\centering
\begin{minipage}{.45\linewidth}
\centering
    \includegraphics[width=0.85\linewidth]{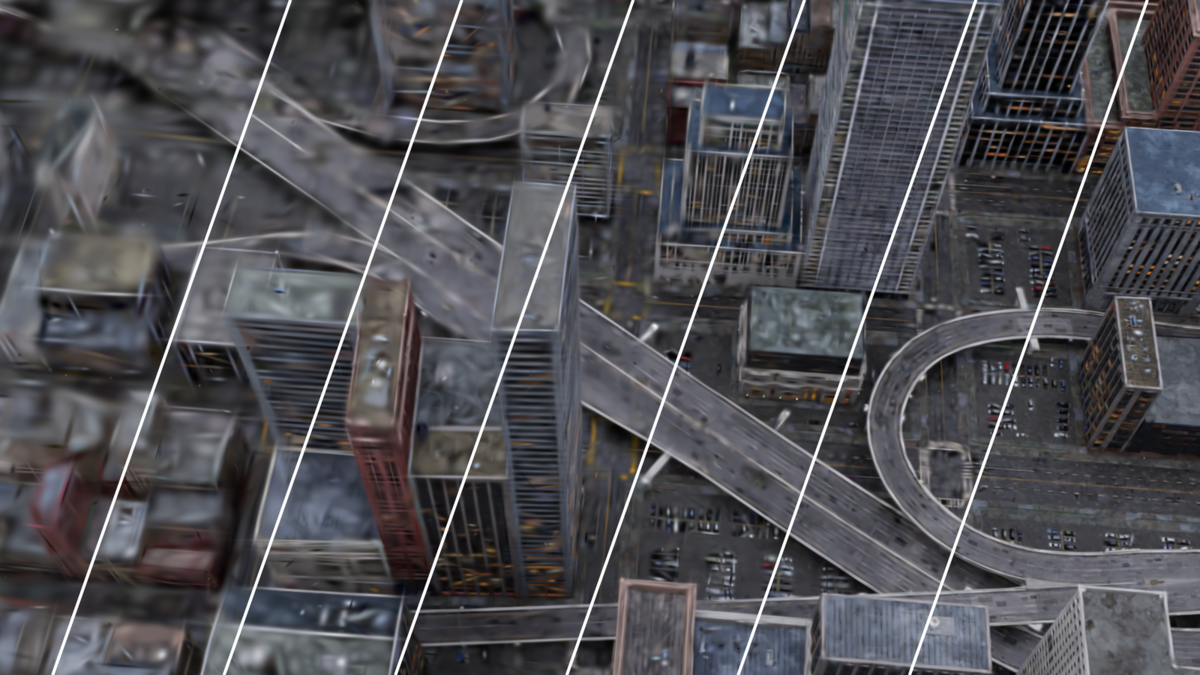}
    \captionof{figure}{Hierarchical SPTs enable smooth transitions between detailed and coarse representation.\label{fig:ContinuousLOD}}
\end{minipage}
\hfill
\end{figure*}



\begin{figure*}
    \setlength{\tabcolsep}{.5pt}%
    \renewcommand{\arraystretch}{.5}
 \resizebox{.95\linewidth}{!}{
    \centering
    \begin{tabular}{ccccccccc}
        && GT & Ours & CityGS & H-3DGS & OctreeGS & HorizonGS  & CLM-GS \\
        \multirow{3}{*}{\rotatebox{90}{MC-small-city+}} &&
        \includegraphics[width=0.16\linewidth]{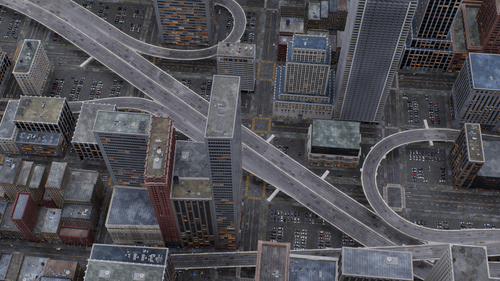} &
        \includegraphics[width=0.16\linewidth]{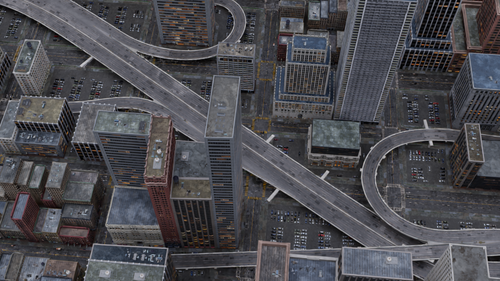} &
        \includegraphics[width=0.16\linewidth]{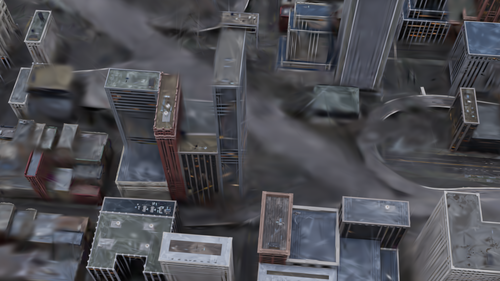} &
        \includegraphics[width=0.16\linewidth]{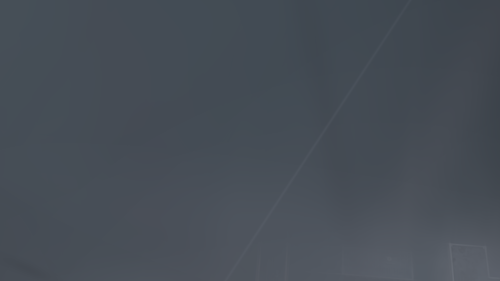} &
        \includegraphics[width=0.16\linewidth]{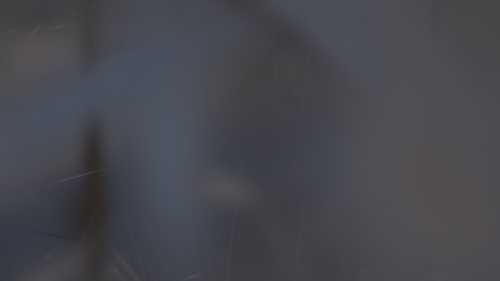} &
        \includegraphics[width=0.16\linewidth]{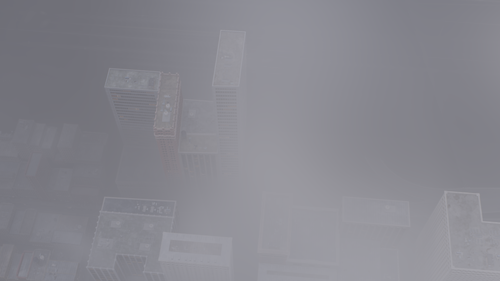} & 
        \includegraphics[width=0.16\linewidth]{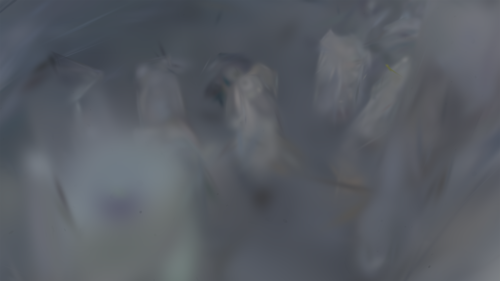}\\
        && 
        \includegraphics[width=0.16\linewidth]{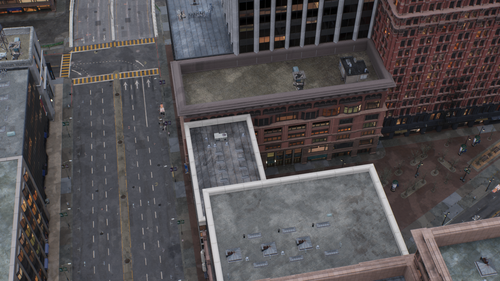} &
        \includegraphics[width=0.16\linewidth]{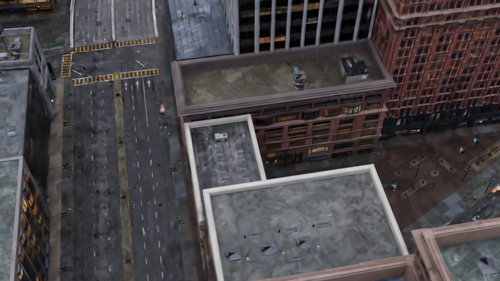} & 
        \includegraphics[width=0.16\linewidth]{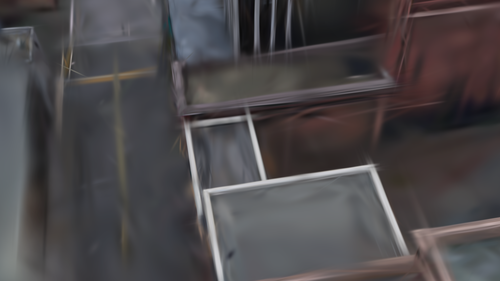} &
        \includegraphics[width=0.16\linewidth]{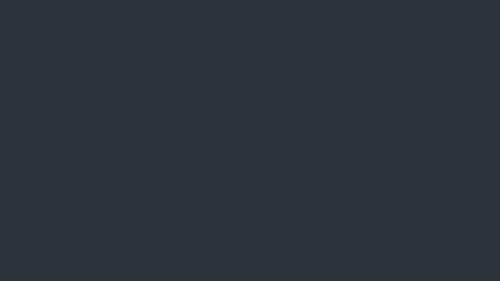} &
        \includegraphics[width=0.16\linewidth]{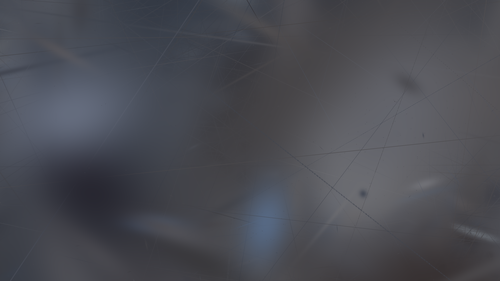} &
        \includegraphics[width=0.16\linewidth]{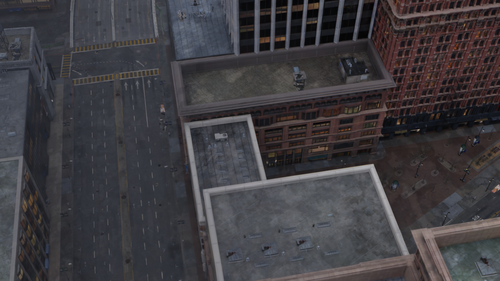} & 
        \includegraphics[width=0.16\linewidth]{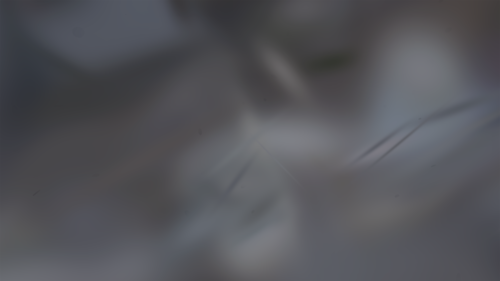}\\
        && 
        \includegraphics[width=0.12\linewidth]{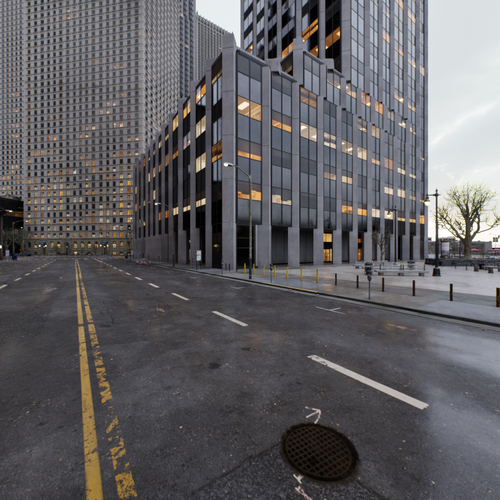} &
        \includegraphics[width=0.12\linewidth]{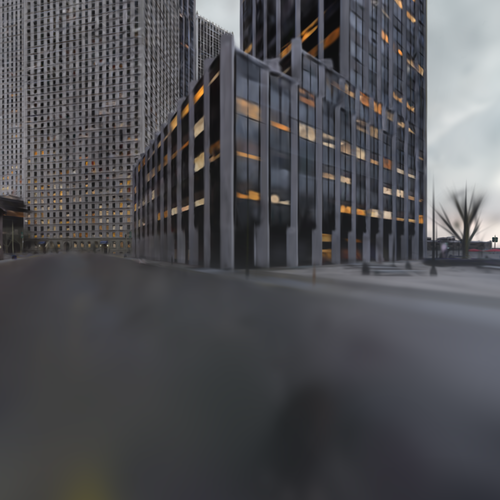} & 
        \includegraphics[width=0.12\linewidth]{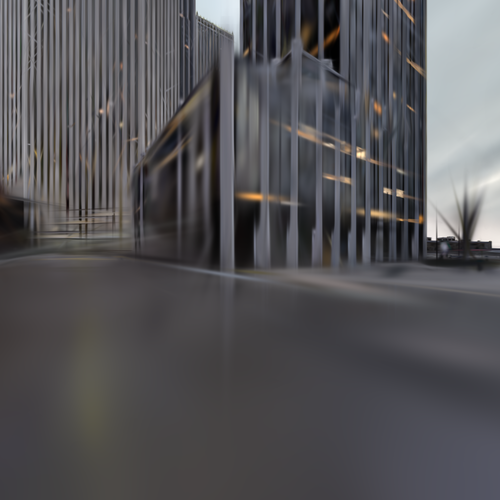} &
        \includegraphics[width=0.12\linewidth]{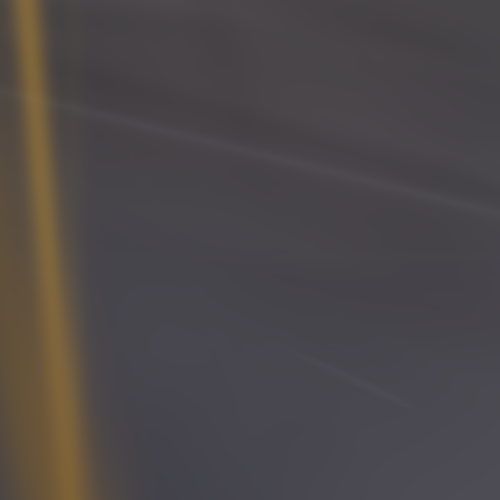} &
        \includegraphics[width=0.12\linewidth]{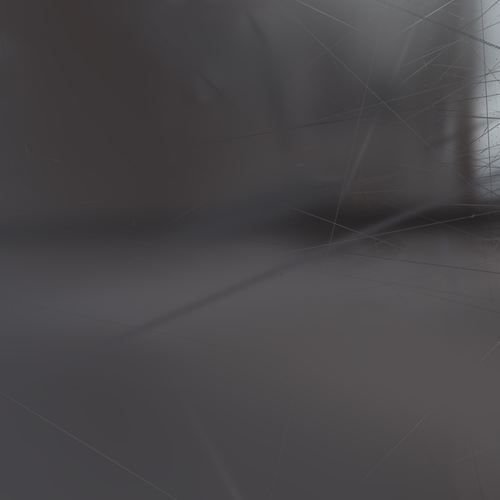} &
        \includegraphics[width=0.12\linewidth]{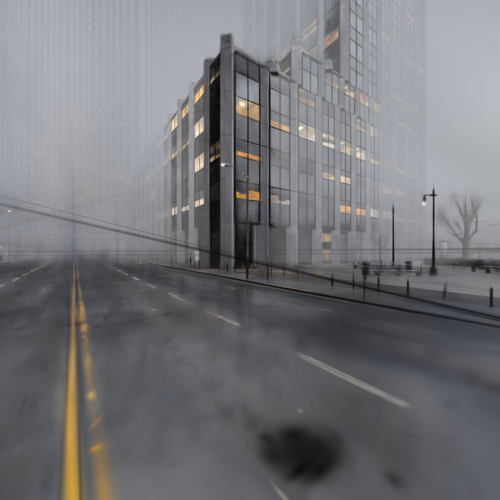} &
        \includegraphics[width=0.12\linewidth]{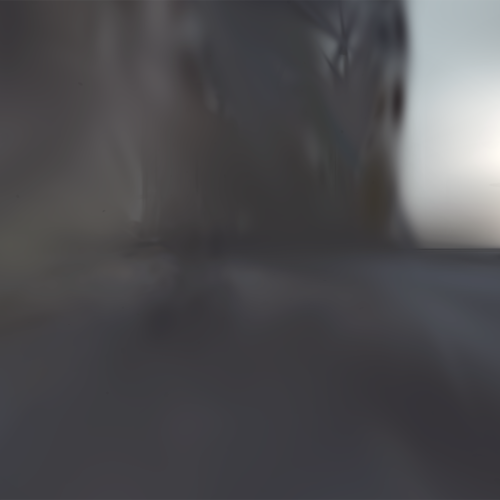}\\

        \multirow{2}{*}[.5em]{\rotatebox{90}{Campus}} &&
        \includegraphics[width=0.16\linewidth]{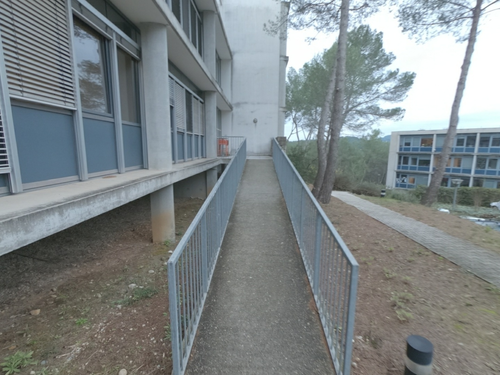} &
        \includegraphics[width=0.16\linewidth]{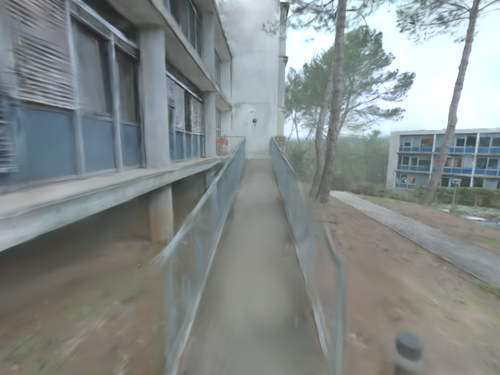} & 
        \includegraphics[width=0.16\linewidth]{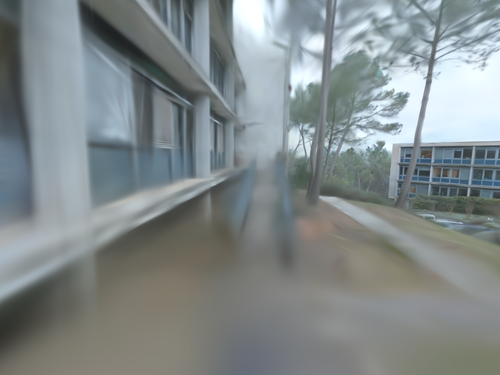} &
        \includegraphics[width=0.16\linewidth]{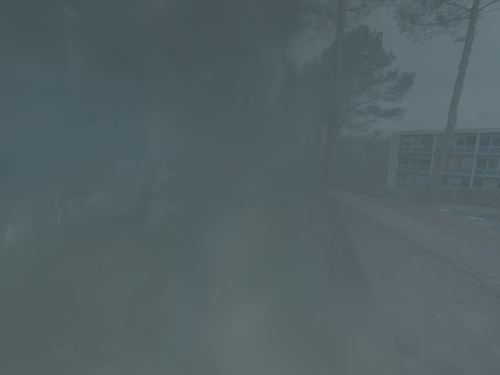} &
        \includegraphics[width=0.16\linewidth]{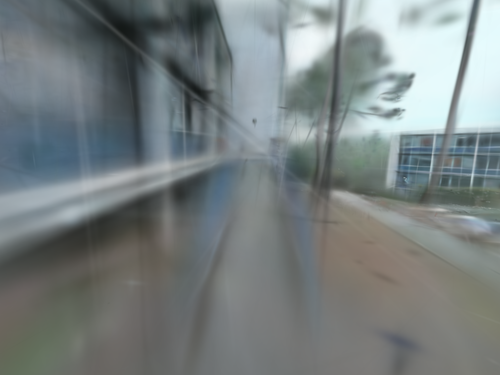} &
        \includegraphics[width=0.16\linewidth]{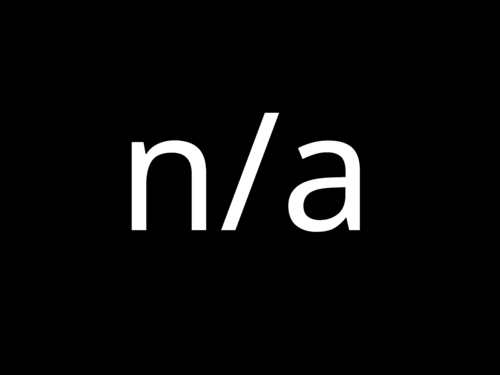} &
        \includegraphics[width=0.16\linewidth]{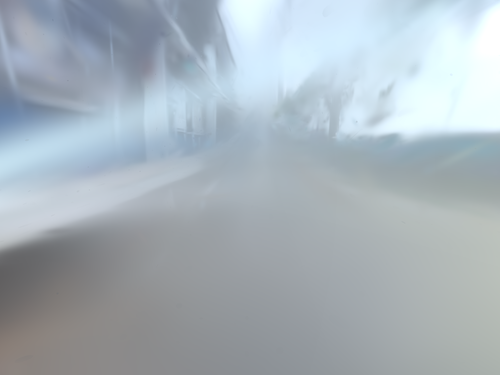} \\
        &&
        \includegraphics[width=0.16\linewidth]{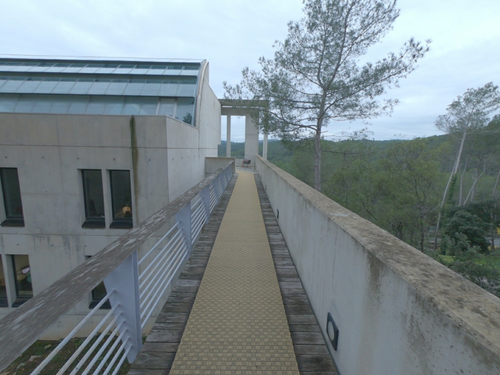} &
        \includegraphics[width=0.16\linewidth]{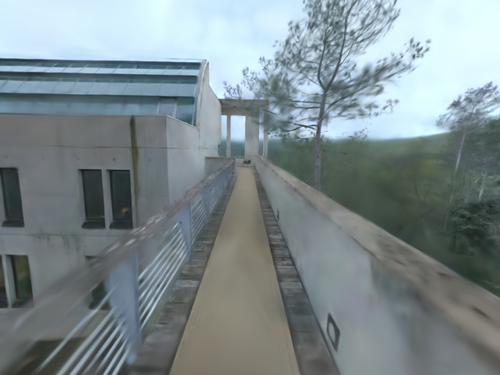} & 
        \includegraphics[width=0.16\linewidth]{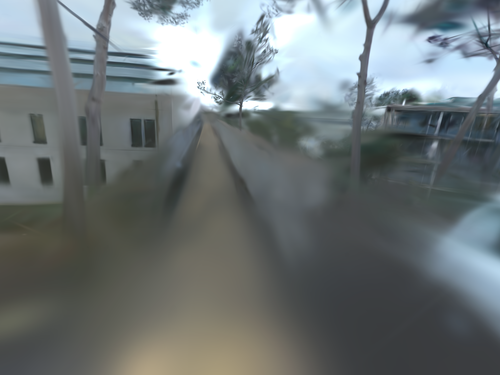} &
        \includegraphics[width=0.16\linewidth]{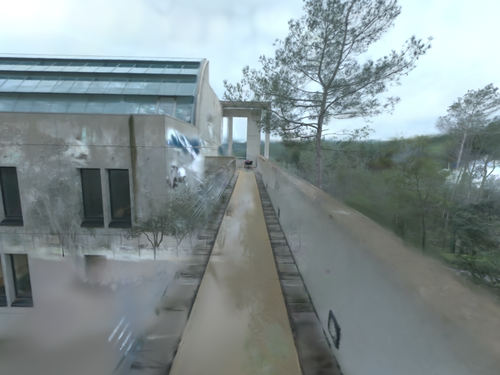} &
        \includegraphics[width=0.16\linewidth]{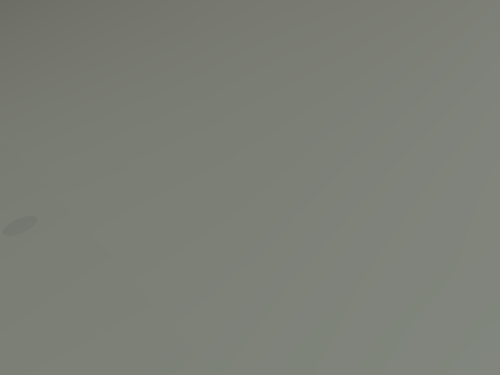} &
        \includegraphics[width=0.16\linewidth]{imgs/Comparison/NA.png} &
        \includegraphics[width=0.16\linewidth]{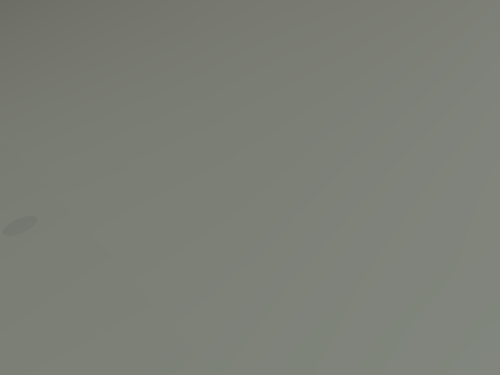} \\
        
        \multirow{2}{*}[.5em]{\rotatebox{90}{Uni10k}} &&
        \includegraphics[width=0.16\linewidth]{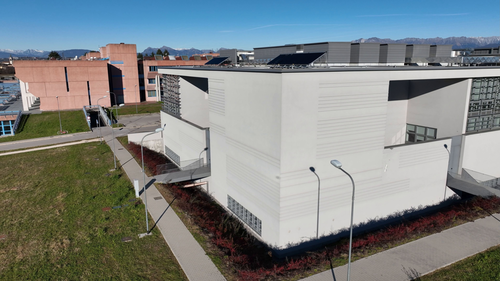} &
        \includegraphics[width=0.16\linewidth]{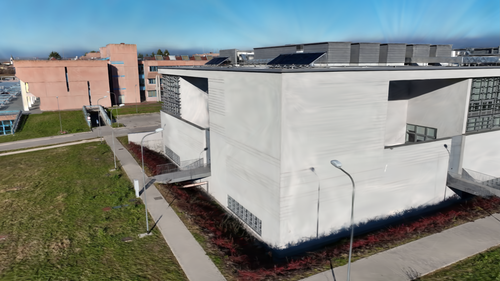} & 
        \includegraphics[width=0.16\linewidth]{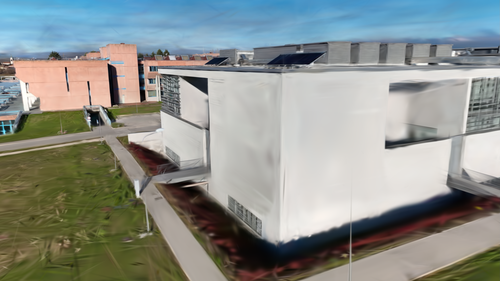} &
        \includegraphics[width=0.16\linewidth]{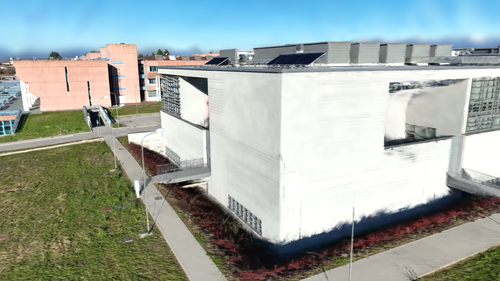} &
        \includegraphics[width=0.16\linewidth]{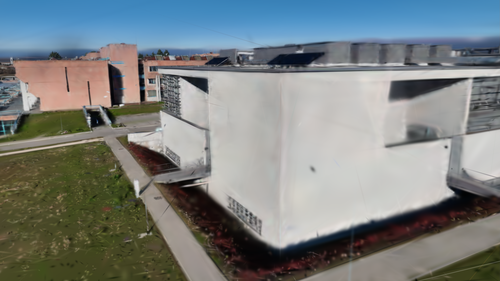} &
       \includegraphics[width=0.16\linewidth]{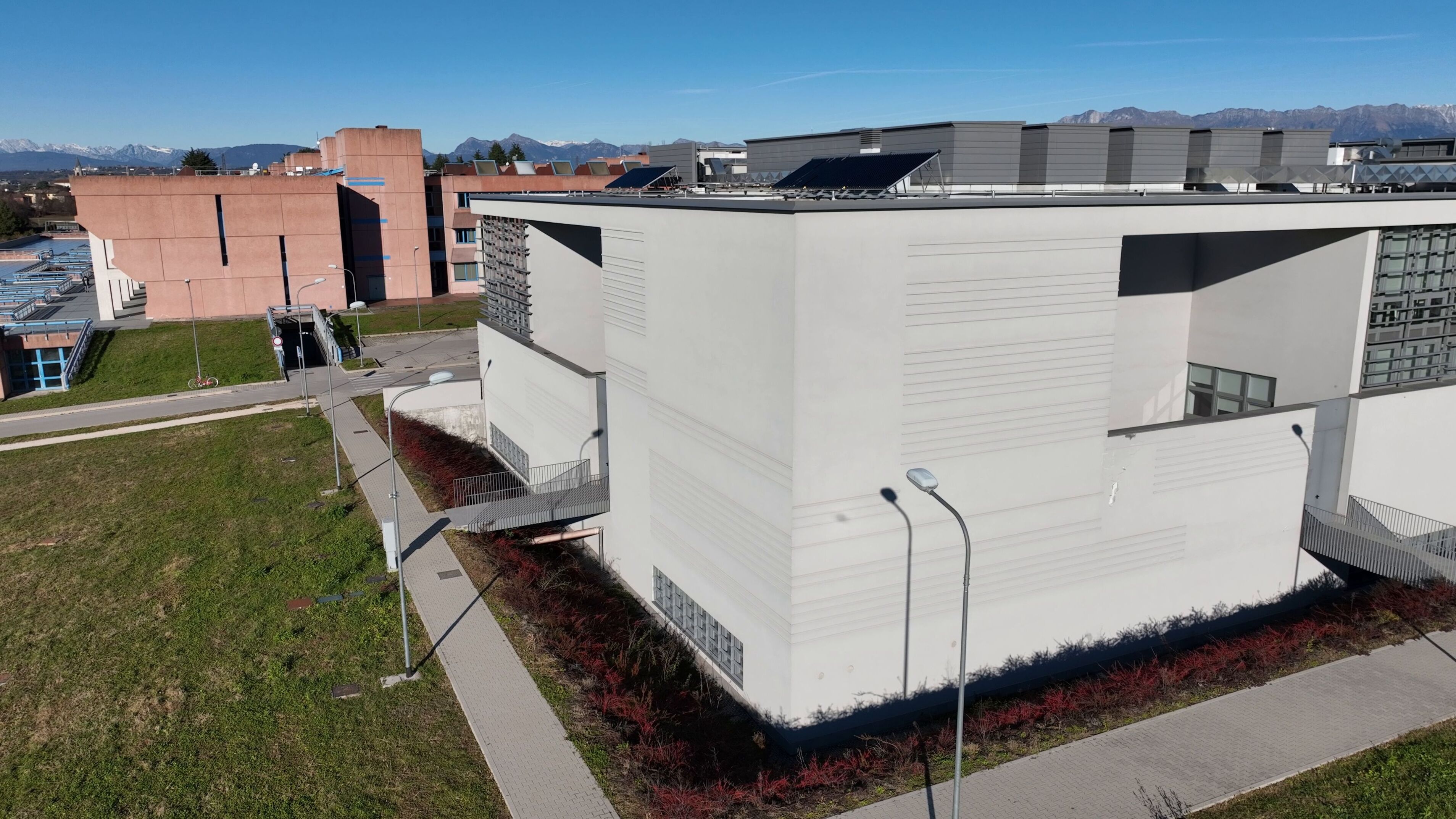} &
        \includegraphics[width=0.16\linewidth]{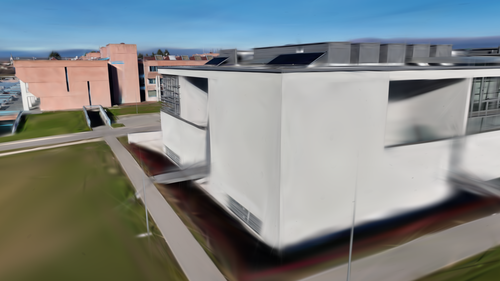}\\
        &&
        \includegraphics[width=0.16\linewidth]{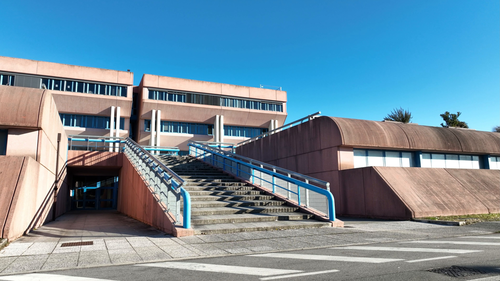} &
        \includegraphics[width=0.16\linewidth]{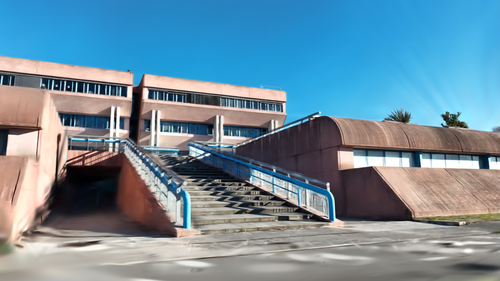} & 
        \includegraphics[width=0.16\linewidth]{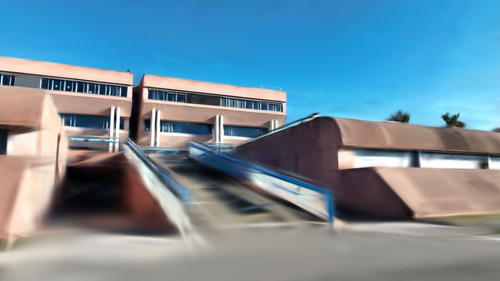} &
        \includegraphics[width=0.16\linewidth]{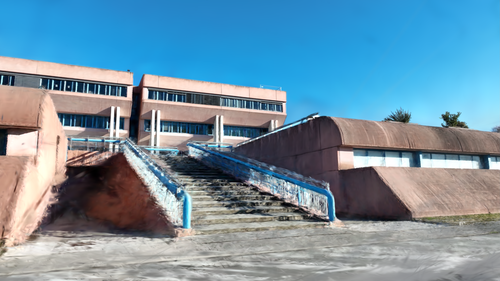} &
        \includegraphics[width=0.16\linewidth]{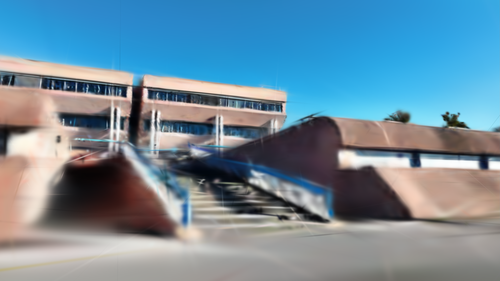} &
        \includegraphics[width=0.16\linewidth]{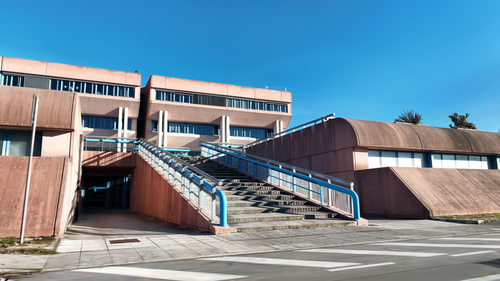} &
        \includegraphics[width=0.16\linewidth]{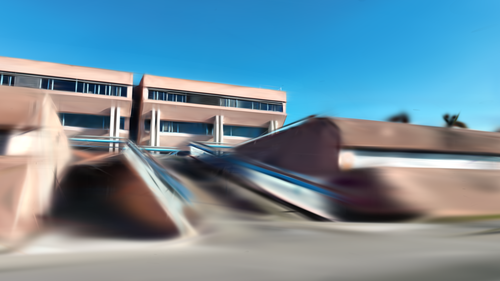} 
    \end{tabular}
     }
    \caption{Qualitative comparison of our method and SOTA methods \citep{CityGaussian, Hierarchical, OctreeGS, CLM} on the MC-small-city+, Campus and Uni10k scenes.}
    \label{fig:Quality}
\end{figure*}

\begin{figure*}[h]
\begin{minipage}{.72\linewidth}   
    \centering
    \begin{subfigure}[t]{0.32\linewidth}
        \includegraphics[width=0.95\linewidth]{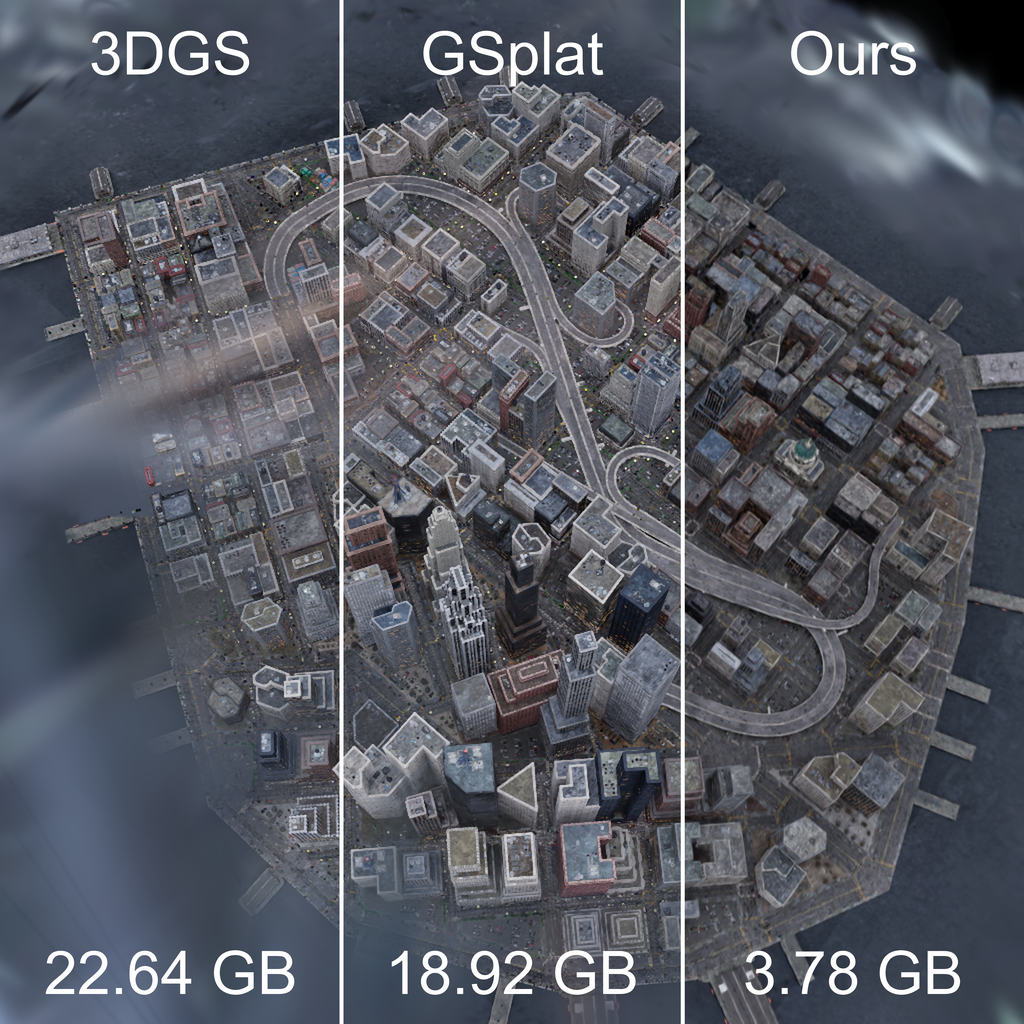}  
        \caption{Overview}
        \label{top}
    \end{subfigure}
    \begin{subfigure}[t]{0.32\linewidth}
        \includegraphics[width=0.95\linewidth]{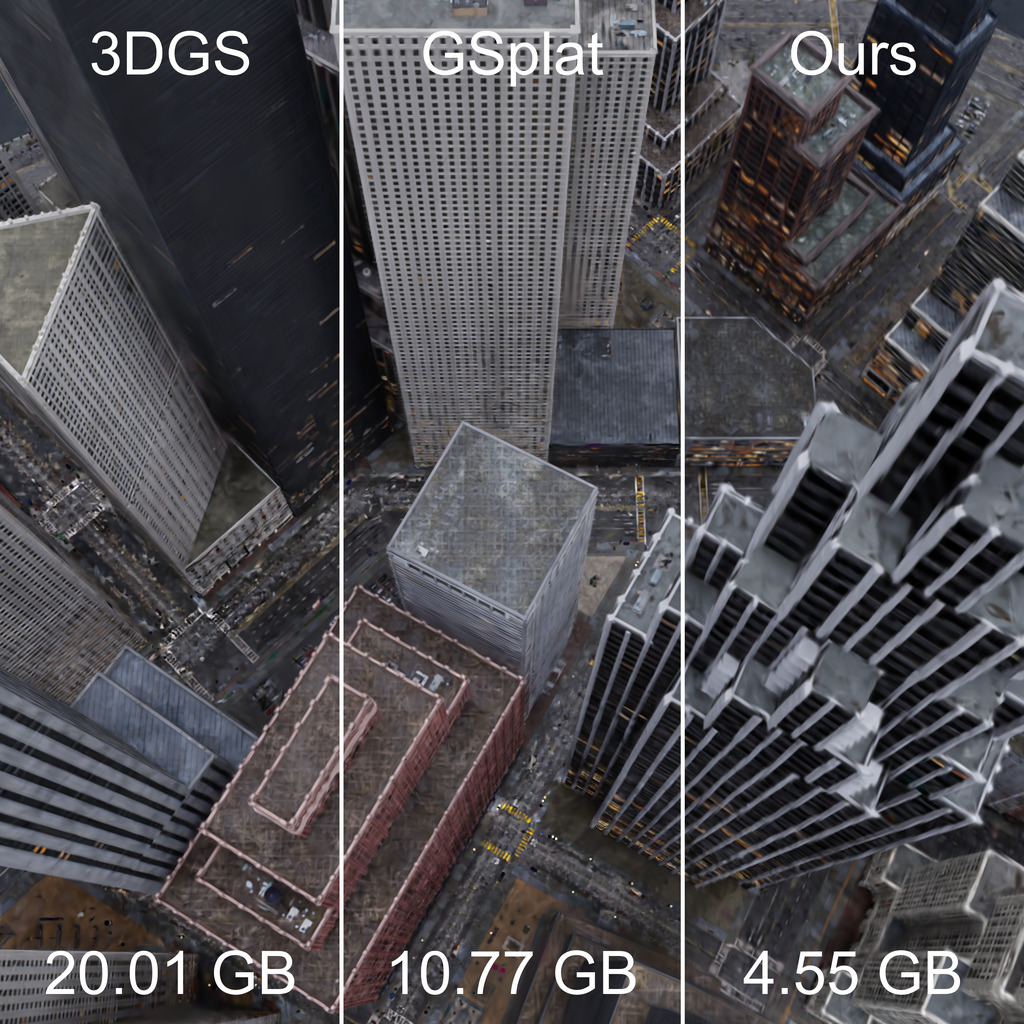}
        \caption{Aerial View}
        \label{bird}
    \end{subfigure}
    \begin{subfigure}[t]{0.32\linewidth}
        \includegraphics[width=0.95\linewidth]{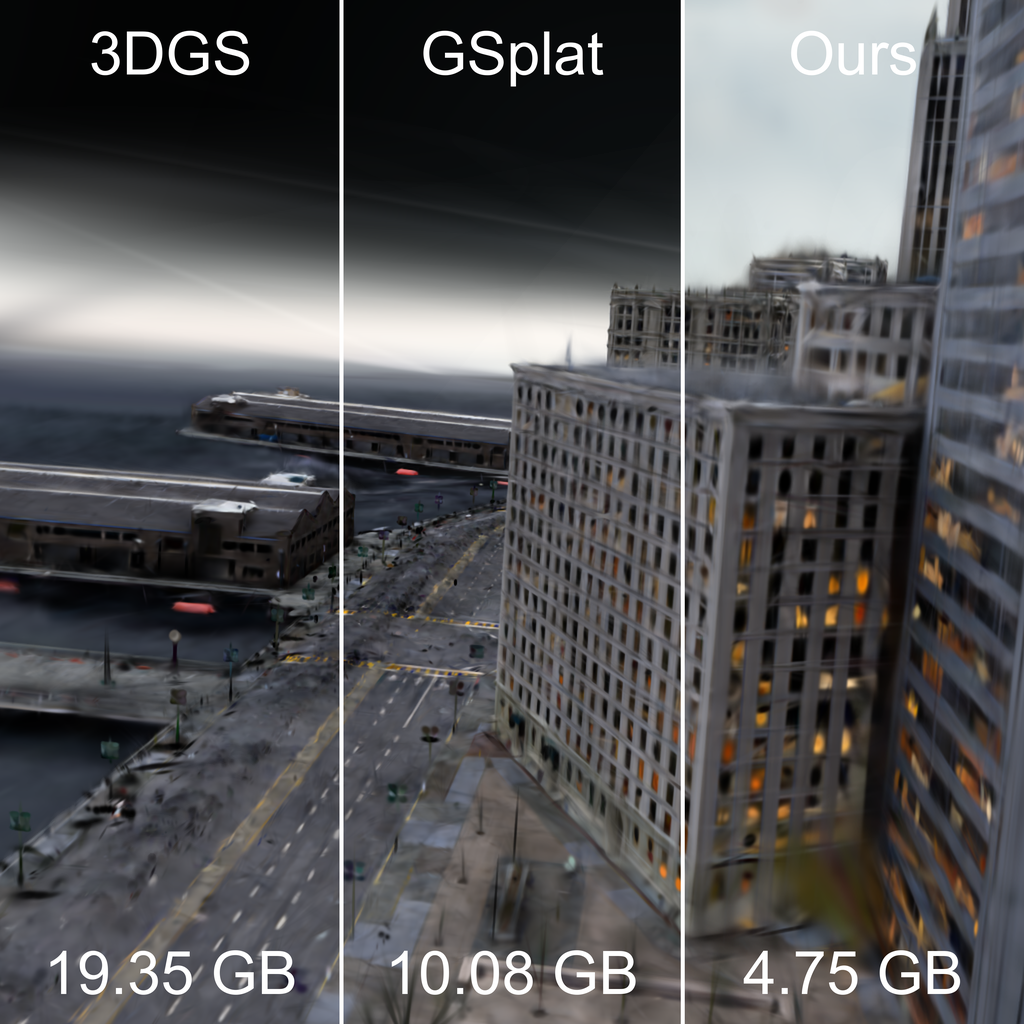}
        \caption{Street View}
        \label{side}
    \end{subfigure}
    \caption{Comparison to standard 3DGS~\citep{3DGS} and gsplat~\citep{Gsplat} in terms of rendering quality and VRAM usage on the MC-small-city+ scene.}
    \label{fig:RenderComp}
\end{minipage}
\hfill
\begin{minipage}{.25\linewidth}
    \centering
    \includegraphics[width=0.9\linewidth]{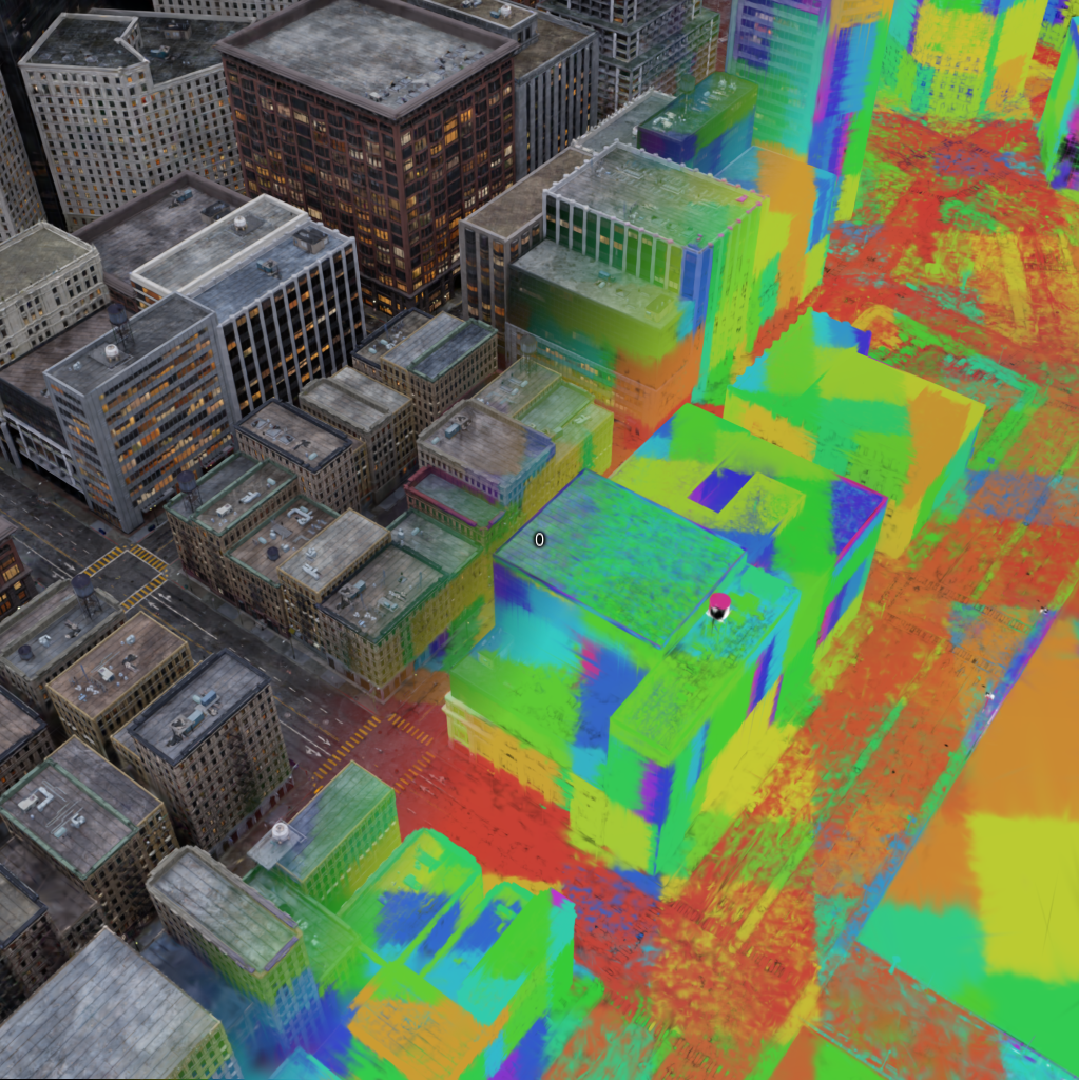}
    \vspace{-8pt}
    \captionof{figure}{SPTs for a frame of MatrixCity rendered in different colors.
    \label{fig:SPTs}}
    \vspace{5pt}
\end{minipage}
\end{figure*}

\FloatBarrier
\appendix
\section{Supplementary Material}

Our full supplementary material consists of:
\begin{enumerate}
    \item This document, which includes additional ablation studies, implementation details and experiments, which we could not include in the main paper due to page restrictions
    \item A short \textbf{video presentation} about this manuscript
    \item An \texttt{html}-page with more qualitative comparisons and short videos
    \item Our full source code contained in the \texttt{Code} directory
    \item Training configurations detailing all hyperparameters for our and baseline methods on each experiment 
\end{enumerate}

\subsection{Additional Ablations}
We present additional experiments to demonstrate the effectiveness of each component of our method.

\paragraph{View Selection}
To verify the effectiveness of caching and view selection, we conduct an experiment by training the MC-smaller-city+ dataset (100k iterations, up to 30M Gaussians) with and without view selection and caching. The results in Table \ref{tab:ViewSelect} show that while quality metrics are barely affected (with a maximal difference of 0.006 PSNR), the number of Gaussians that need to be loaded into VRAM decrease by a factor of almost $20\times$ with view selection and caching enabled. 

\begin{table}[t]
    \centering
        \caption{Ablation results on the MC-smaller-city+ dataset. \#Loaded refers to the average number of Gaussians loaded from RAM over 100\,000 iterations.}
\begin{tabular}{lrrrr}
\toprule
 & PSNR\textsuperscript{$\uparrow$} & SSIM\textsuperscript{$\uparrow$} & LPIPS\textsuperscript{$\downarrow$}  &  \#Loaded  \\
\midrule
 w/o caching        & 21.831 & 0.701 & 0.425  & 1\,629\,187 \\
 w/ caching         & 21.829 & 0.700 & 0.429  & 290\,572 \\
 w/ view selection  & 21.825 & 0.697 & 0.432  & 89\,451 \\
\bottomrule
\end{tabular}

\label{tab:ViewSelect}
\end{table}

\paragraph{Effectiveness of LoD}
To evaluate the effectiveness of our level-of-detail system, we choose 5 random images from each dataset and compare the resulting quality metrics to the number of Gaussians rendered at 50 different levels of details. The results are visualized in Figure \ref{fig:LODEVAL} in addition to average FPS for rendering the view 10 times with pre-warmed cache on an H200 GPU. As expected, FPS scales closely with the inverse of the number of Gaussians. In the vast majority of cases, increasing the LoD level consistently improves rendering quality. Overall, we only notice a small dip in quality when reducing the Gaussian count by 50\%, verifying the effectiveness of our LoD structure. Further, using continuous instead of discrete levels of detail leads to the smoothness of the curves in Figure \ref{fig:LODEVAL}, indicating a smooth transition between LoD levels, which reduces popping.

\begin{figure}[h]
    \centering
    \begin{subfigure}{0.45\linewidth}
        \includegraphics[width=0.95\linewidth]{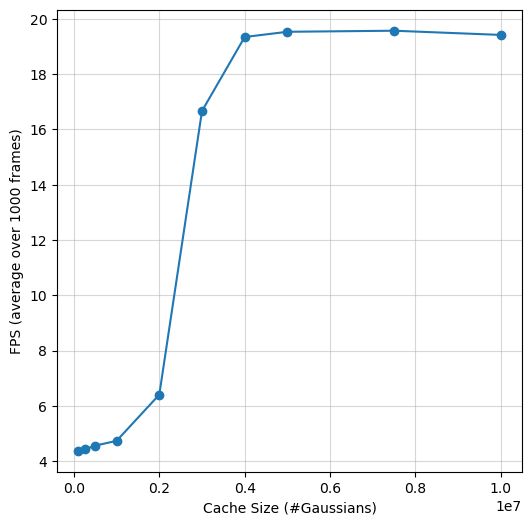}
        \caption{MC-small-city+}
    \end{subfigure}
    \begin{subfigure}{0.45\linewidth}

    \includegraphics[width=0.95\linewidth]{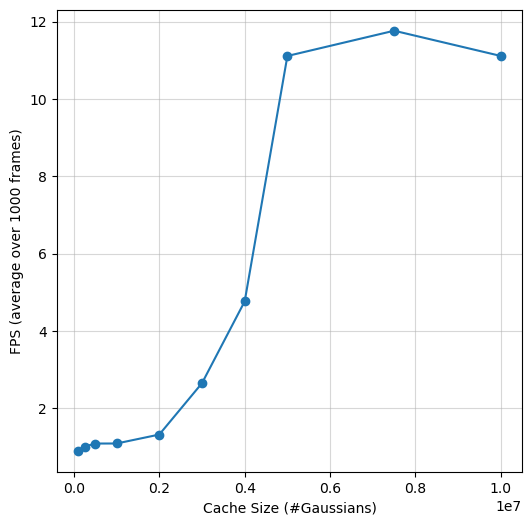}
        
        \caption{Campus}
    \end{subfigure}
    \caption{Average FPS for rendering 1000 frames of the camera paths for the MC-small-city+ and Campus scenes with various cache sizes.}
    \label{fig:CacheSizeFPS}
\end{figure}

\begin{figure*}[h]
    \centering
    \begin{subfigure}{0.32\linewidth}
    \includegraphics[width=0.95\linewidth]{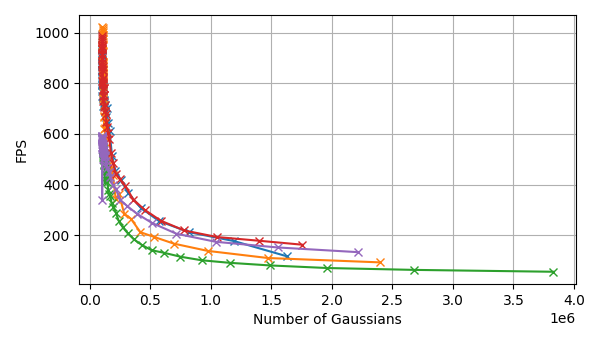}
        \includegraphics[width=0.95\linewidth]{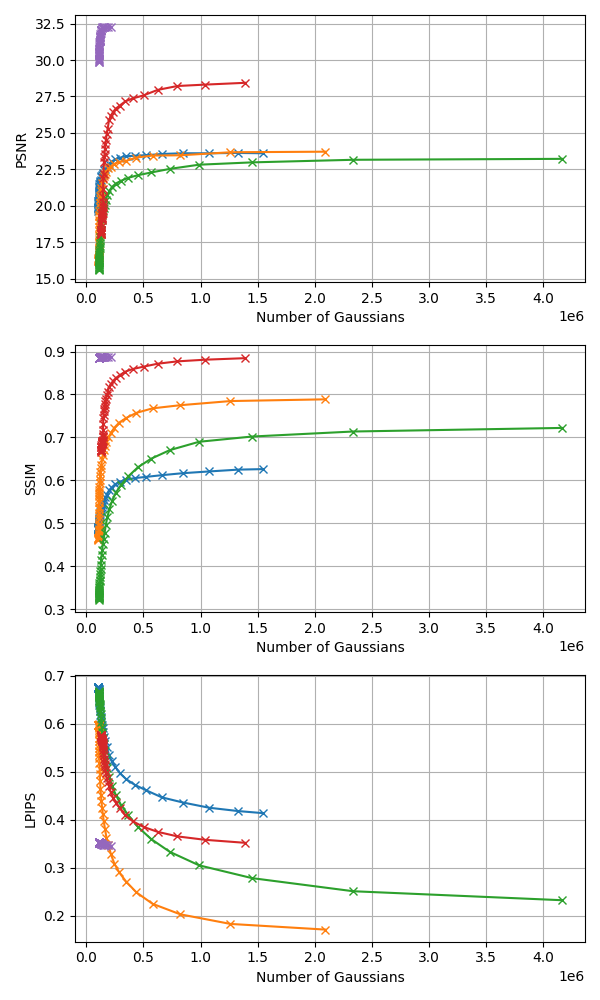}
        \caption{MC-small-city+ aerial.}
    \end{subfigure}
    \begin{subfigure}{0.32\linewidth}
    \includegraphics[width=0.95\linewidth]{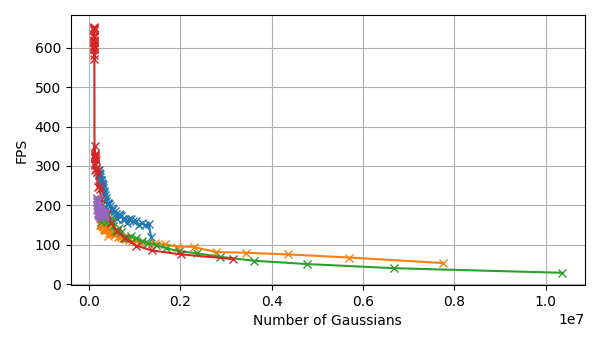}
        \includegraphics[width=0.95\linewidth]{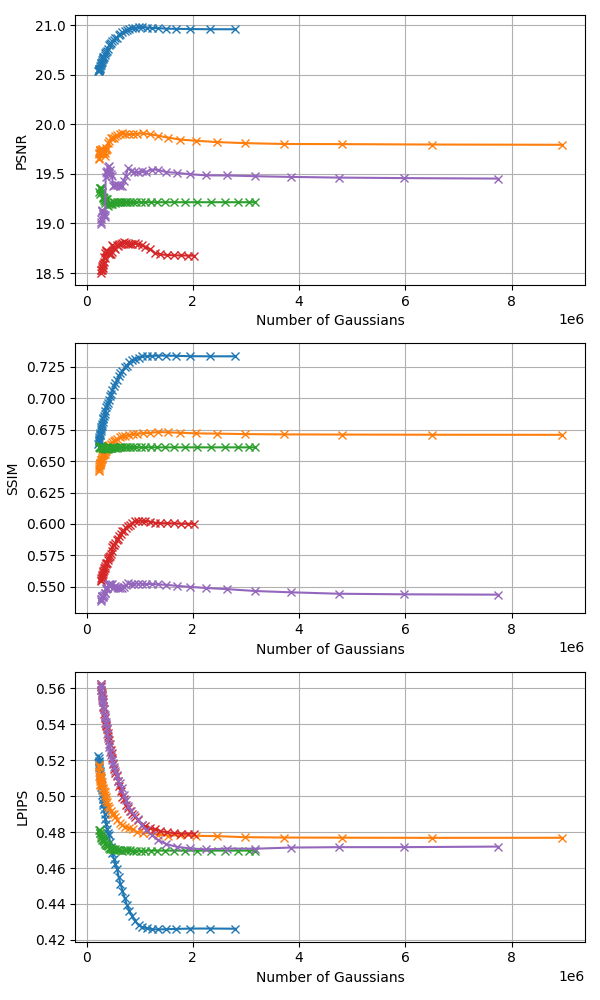}
        \caption{MC-small-city+ street.}
    \end{subfigure}
    \begin{subfigure}{0.32\linewidth}
    \includegraphics[width=0.95\linewidth]{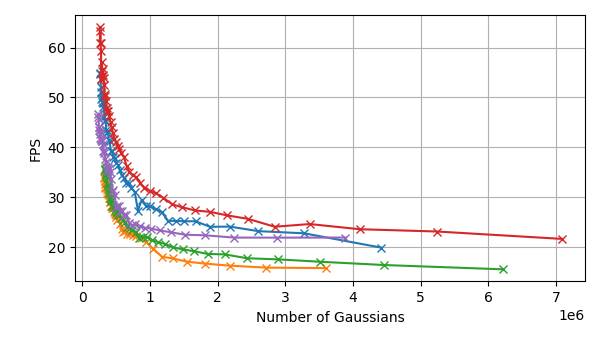}
    \includegraphics[width=0.95\linewidth]{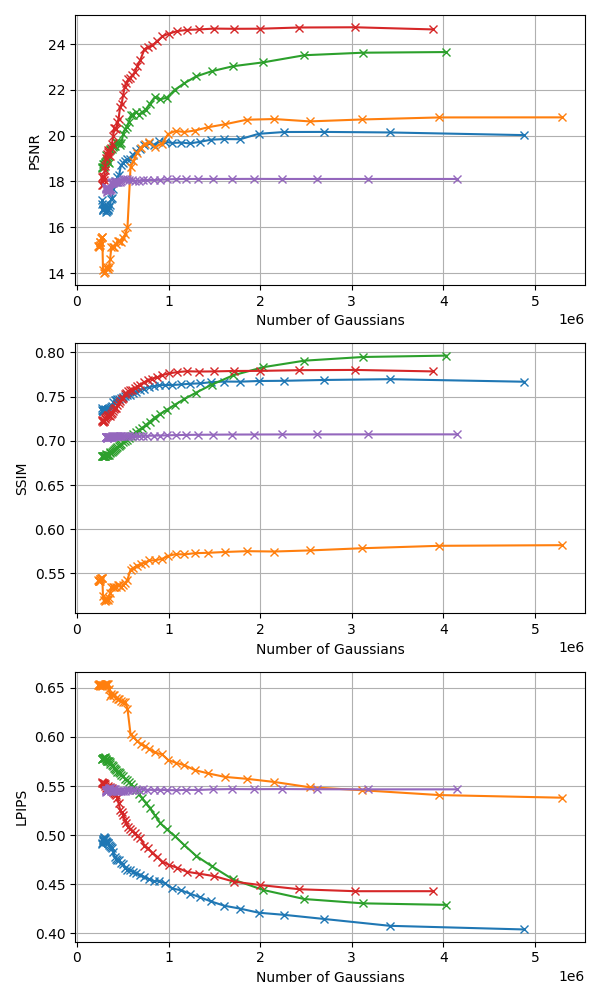}
        
        \caption{Campus.}
    \end{subfigure}
    \caption{Qualitative evaluation of our LoD system for 5 randomly chosen views and 50 different levels of detail.}
    \label{fig:LODEVAL}
\end{figure*}

\paragraph{Effectiveness of Caching}
In Figure \ref{fig:CacheSizeFPS}, we provide sensitivity curves for cache size vs. FPS during rendering. In each iteration up to \emph{cache size} Gaussians from the previous frames may be reused. The results once again demonstrate the importance of the caching system to rendering efficiency. The graphs show that performance increases steeply with increasing cache size until it exceeds the typical number of Gaussians required per iteration and starts to plateau.


\subsection{Initialization and Training Details}
\label{App:Initialization}
Following \citet{Hierarchical}, we initialize the Gaussian model from a sparse point cloud, augmented with skybox points. This initial representation is small enough to fully reside in GPU memory and is trained for $100$k iterations without densification. The goal of this phase is to establish a stable global scene structure before constructing the hierarchy. After this initial optimization, we build a binary Gaussian hierarchy, where the trained Gaussians act as leaf nodes and the parent nodes represent merged approximations of their two children.

The Gaussian properties--including base color, SH coefficients, position, and covariance--are optimized using the standard loss propagation introduced by \citet{3DGS}.
To fulfill our tight memory requirements, we update only the parameters of the cut set chosen for each training view.
Thus, gradients may, in general, be propagated to Gaussians in the middle of the hierarchy. Assuming good coverage and diversity of training views, these updates will diffuse over the entire hierarchy over the coarse of training, leading to smooth transitions from the highest to the lowest LoD. 
\color{black}

\subsection{Dataset Details and Further Results}
\label{App:Datasets}
We include more details on the datasets used in the evaluation and provide additional results for smaller-scale scenes.

\paragraph{MC-small-city+} 
For our main benchmark, {MC-small-city+, we aggregate 33\,006 street-view images and 7\,672 aerial views from the small-city scene of the MatrixCity dataset \citep{MatrixCityDataset} and generate 533 additional high-altitude views. We evaluate reconstruction quality on a separate set of 4\,228 test views, according to the test split provided by the dataset. The scene is extremely challenging due to its enormous scale, sparse views and wide variation in scale: As such, we also construct a subset of the entire dataset (15.1k images, covering about a third of the area), denoted as MC-smaller-city+ for baselines that are not able to reconstruct the full dataset. For MC-small-city+, we use the camera poses provided by the MatrixCity dataset and convert them to the COLMAP format for Gaussian splatting.
We merge the provided street and aerial sparse point clouds and randomly downsample them by a factor of $5$ to get more realistic initialization conditions. We do not make use of the ground truth depth images provided by the dataset, as we consider this to be an unrealistic advantage. Figure \ref{fig:BigCityScaleDataset} shows the scene along with the camera distribution.
While some methods have successfully reconstructed only aerial \citep{CityGaussian, CityGaussianV2, OctreeGS} or only street-level views \citep{CityGaussianV2, NerfXL}, training a model that holds up to scrutiny from both perspectives presents a particular challenge.
Training on close and far views simultaneously--without a proper LoD system like ours in place--significantly degrades visual quality for both, as noted by \citet{HorizonGS, CrossView}.
Moreover, such a scenario greatly complicates partitioning the scene into independent chunks.

\paragraph{Why are the quality metrics on MC-small-city+ so low?}
When comparing our evaluation on the MatrixCity dataset with those in other works like \citet{CityGaussian, CLM, HorizonGS, OctreeGS}, it might seem surprising that they achieve significantly lower quality metrics on our constructed MC-small-city+ scene. There are several reasons for this:
\begin{enumerate}
    \item These works only evaluate on the aerial views, which make up less than 20\% of the scene images and are generally easier to reconstruct. 
    \item Reconstructing street, aerial and the newly generated high-aerial views in the same scene is much more difficult than reconstructing just aerial views.

    \item Some competing works \citep{CityGaussian, CLM} use the ground-truth depth maps or dense point clouds from the MatrixCity dataset. We would consider this to be an unrealistic advantage, since they contain almost the entire ground truth geometry of the scene. \citet{CLM} even starts the reconstruction from the full dense point cloud and never densifies, whereas our evaluation starts from the same point cloud downsampled by a factor of 5.
\end{enumerate}

\begin{figure*}[h]
    \centering
    \includegraphics[width=0.7\linewidth]{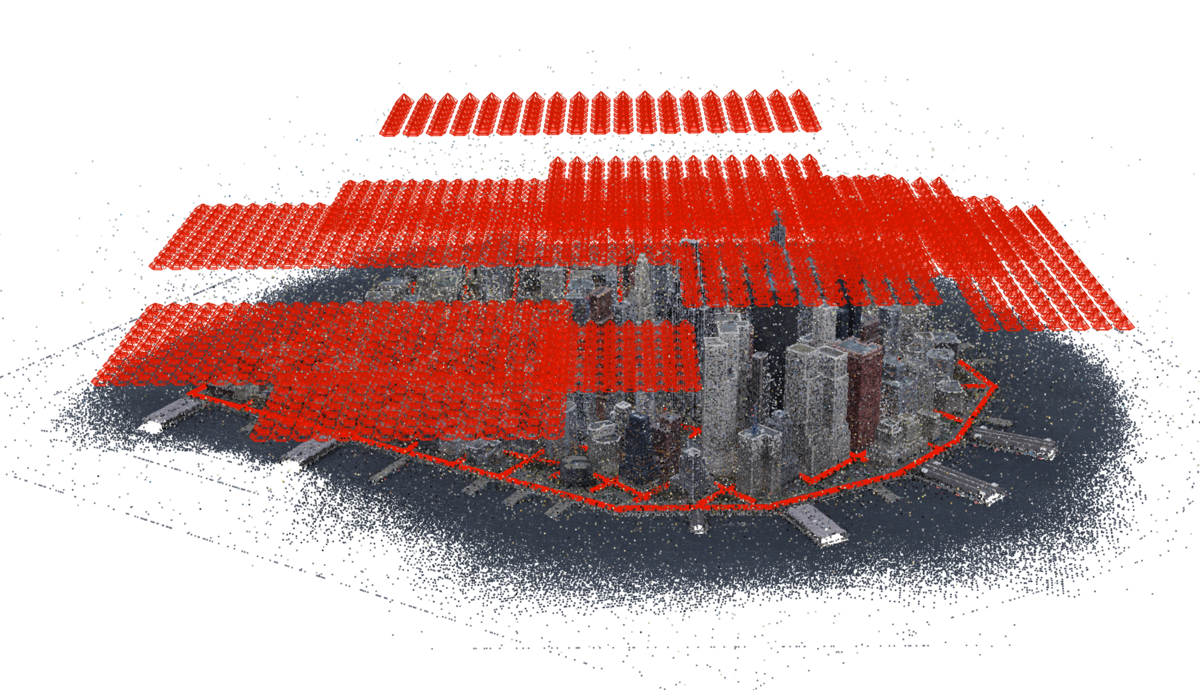}
    \caption{The MC-small-city+ dataset spans hundreds of buildings, which are supervised by tens of thousands of views, varying widely in scale.}
    \label{fig:BigCityScaleDataset}
\end{figure*}

\paragraph{Uni10K}
In order to address the gap of real-world large-scale datasets that are captured across multiple scales, we present Uni10k. The scene consists of an outdoor campus of approximately $100\,000 \text{ m}^2$, captured from both ground-level and aerial perspectives. Standard reconstruction via COLMAP at this scale would typically require weeks of computation due to the complexity of the image-matching and mapping stages. To mitigate this, we leverage spatial and temporal priors in conjunction with a coarse-to-fine scheme. Specifically, we utilize GPS data to reduce the matching complexity from quadratic to near-linear by limiting image comparisons to a predefined spatial radius. Regarding temporal priors, since frames are sampled from video sequences, we initially reconstruct a baseline model starting from a set of frames uniformly sampled every second of video. Then, we densify the camera coverage by incrementally registering, triangulating, and refining new images using only local BA. The process concludes with several rounds of Global bundle adjustment. The final reconstruction comprises more than $10\,000$ images at 4K resolution and ~6.2M sparse points, with an overall MRE of approximately $0.63$ pixels. We train on the full resolution images and hold back every 8th frame alphabetically for the test set. Figure \ref{fig:uniud10k} shows the sparse point cloud and camera distribution.

\begin{figure}[t]
    \centering
    \includegraphics[width=\linewidth]{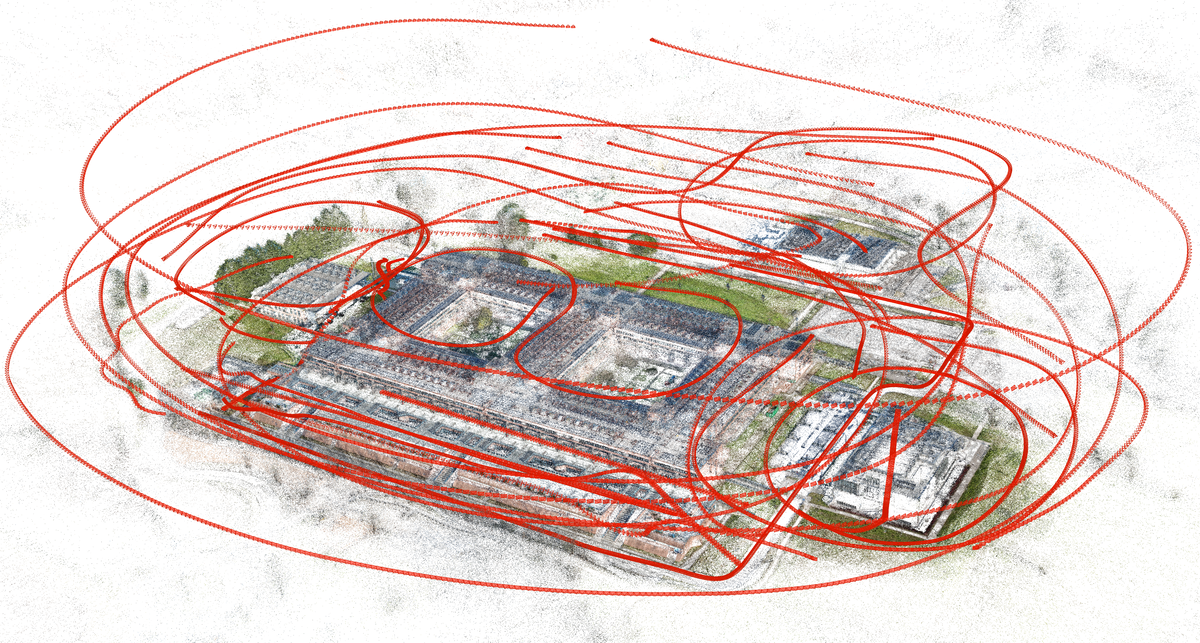}
        \caption{Uni10K's sparse point cloud. The scene is reconstructed from more than $10$ thousand 4k images at widely varying heights, both from aerial and ground perspective.}
    \label{fig:uniud10k}
\end{figure}

\paragraph{Hierarchical 3DGS dataset}
Since H-3DGS \citet{Hierarchical} did not introduce a test split for their dataset, we hold back every $100$th image alphabetically for testing.
We reevaluate H-3DGS for this scene with the new test split, using the camera calibrations and chunk splits provided on their website. In accordance with the instructions, we disable exposure optimization for evaluation. On this dataset only, we use depth supervision identical to \citet{Hierarchical}. The distribution of street views in the Campus scene is visualized in Figure \ref{campus}.\\
\paragraph{Hierarchical 3DGS single-chunk dataset} To demonstrate our ability to reconstruct small scenes, we also evaluate our method on the smaller, single-chunk versions of the Campus and Small City scenes (cf. Table \ref{tab:SingleChunk}). The quality metrics indicate that training on the LoD structure (which would not be required for scenes of this scales) during training only leads to marginal reductions in visual quality. \\
\begin{table}[t]
    \centering
        \caption{
{H-3DGS single chunk view synthesis results.} Results with \textsuperscript{\textdagger} are taken from \cite{Hierarchical}.
}
\resizebox{1.0\linewidth}{!}{
\begin{tabular}{lcccccc}
\toprule
 & \multicolumn{3}{c}{Small City Chunk (1.1k images)} & \multicolumn{3}{c}{Campus Chunk (1.5k images)} \\
\cmidrule(lr){2-4}\cmidrule(lr){5-7}
Method & PSNR\textsuperscript{$\uparrow$} & SSIM\textsuperscript{$\uparrow$} & LPIPS\textsuperscript{$\downarrow$}  & PSNR\textsuperscript{$\uparrow$} & SSIM\textsuperscript{$\uparrow$} & LPIPS\textsuperscript{$\downarrow$}  \\
\midrule
Mip-NeRF 360 \textsuperscript{\textdagger}  & 24.70 & 0.765 & 0.348  & 20.95 & 0.731 & 0.442 \\
INGP-big \textsuperscript{\textdagger}  & 23.47 & 0.715 & 0.426   & 20.37 & 0.700 & 0.476  \\
F2-NeRF-big \textsuperscript{\textdagger}  & 24.53 & 0.762 & 0.342   & 19.25 & 0.681  & 0.478  \\
3DGS\textsuperscript{\textdagger}  & 25.34 & 0.776  & 0.337   & 23.87 & 0.785 & 0.378 \\
H-3DGS\textsuperscript{\textdagger}  & \cellcolor{blue!25} 26.62 & \cellcolor{blue!25} 0.820 & \cellcolor{blue!25} 0.259    & \cellcolor{blue!10} 24.61 & \cellcolor{blue!25} 0.807 & \cellcolor{blue!25} 0.331  \\
Ours   & \cellcolor{blue!10} 25.94 & \cellcolor{blue!10} 0.808 & \cellcolor{blue!10} 0.276 & \cellcolor{blue!25} 24.86 &  \cellcolor{blue!10} 0.796 & \cellcolor{blue!10} 0.360\\
\bottomrule
\end{tabular}
}
\label{tab:SingleChunk}
\end{table}

\paragraph{OccluGaussian Dataset}
The OccluGaussian dataset features three large-scale indoor environments. Unfortunately, the download link for the Gallery scene on the official website leads to a broken zip file. As we have not received an answer from the authors, we were unable to evaluate on this scene. Additionally, some training images that are referenced in the camera pose files were not included in the download of the Canteen scene. We ran the evaluation on all images that were provided using full resolution images.

\paragraph{Mill19 and Urbanscene3D dataset} Table \ref{tab:Mill19} and \ref{tab:UrbanScene3D} show results on the Mill19~\citep{MegaNerf} and UrbanScene3D~\citep{UrbanScene3D} datasets. We use the camera poses and sparse point cloud provided by \citet{CityGaussian}. In accordance with baselines, we downscale all images by a factor of 4. We have included these dataset, because they are widely used in large-scale novel-view synthesis. At the same time, they represents a worst-case scenario for our method, as the regular, same-height aerial views (cf. Figure \ref{rubble}) negate any benefit of our level-of-detail method and can be split into independent chunks trivially.

\begin{table}[h]
    \centering
        \caption{
\textbf{Mill19 novel view synthesis results} Results with \textsuperscript{\textdagger} are taken from \citet{CityGaussian}. Results of \emph{H-3DGS} are taken from \citet{Hierarchical}.
}
\resizebox{.98\linewidth}{!}{
\begin{tabular}{lrrrrrrrr}
\toprule
 & \multicolumn{4}{c}{Rubble (1.6k images)} & \multicolumn{4}{c}{Building (1.9k images)} \\
\cmidrule(lr){2-5}\cmidrule(lr){6-9}
Method & PSNR\textsuperscript{$\uparrow$} & SSIM\textsuperscript{$\uparrow$} & LPIPS\textsuperscript{$\downarrow$} & Size & PSNR\textsuperscript{$\uparrow$} & SSIM\textsuperscript{$\uparrow$} & LPIPS\textsuperscript{$\downarrow$} & Size \\
\midrule
MegaNeRF\textsuperscript{\textdagger}  & 24.06 & 0.553 & 0.516 & n/a  & 20.93 & 0.547 & 0.504 & n/a \\
3DGS\textsuperscript{\textdagger}   & \cellcolor{blue!10} 25.47 & \cellcolor{blue!10} 0.777 & \cellcolor{blue!10} 0.277 & 6.1M  &  20.46 &  0.720 &  0.305 & 6.4M\\ 
CityGaussian\textsuperscript{\textdagger} & \cellcolor{blue!25} 25.77 & \cellcolor{blue!25} 0.813 & \cellcolor{blue!25} 0.228 & 9.7M  & \cellcolor{blue!25} 21.55 & \cellcolor{blue!25} 0.778 & \cellcolor{blue!25} 0.246 & 13.2M\\
H-3DGS  & 24.64 & 0.755 & 0.284  & n/a  &  \cellcolor{blue!10} 21.52 & \cellcolor{blue!10} 0.723 & \cellcolor{blue!10} 0.297  & n/a\\
Ours & 23.75 & 0.704 & 0.330 & 25M & 20.67 & 0.682 & 0.318 & 25M \\
\bottomrule
\end{tabular}
}\label{tab:Mill19}
\end{table}

\begin{table}[h]
    \centering
        \caption{
\textbf{UrbanScene3D novel view synthesis Results}. Results with \textsuperscript{\textdagger} are taken from \citet{CityGaussian}.}
\resizebox{.98\linewidth}{!}{
\begin{tabular}{lrrrrrrrr}
\toprule
 & \multicolumn{3}{c}{Residence (2.5k images)} & \multicolumn{3}{c}{Sci-Art (3k images)} \\
\cmidrule(lr){2-4}\cmidrule(lr){5-7}
Method & PSNR\textsuperscript{$\uparrow$} & SSIM\textsuperscript{$\uparrow$} & LPIPS\textsuperscript{$\downarrow$}  & PSNR\textsuperscript{$\uparrow$} & SSIM\textsuperscript{$\uparrow$} & LPIPS\textsuperscript{$\downarrow$}  \\
\midrule
CityGaussian\textsuperscript{\textdagger}  & \cellcolor{blue!25}21.90 & \cellcolor{blue!25}0.805 & \cellcolor{blue!25}0.217  &  \cellcolor{blue!25}21.34 & \cellcolor{blue!25}0.833 &  \cellcolor{blue!25}0.232 \\
3DGS\textsuperscript{\textdagger}  & \cellcolor{blue!10}21.44 &\cellcolor{blue!10} 0.791 &\cellcolor{blue!10} 0.236  & \cellcolor{blue!10}21.05 & \cellcolor{blue!10}0.830 &\cellcolor{blue!10} 0.242 \\
H-3DGS &    17.52 & 0.416 & 0.391 & 18.42 & 0.652 & 0.340 \\
OctreeGS & 18.76 & 0.523 & 0.539 & 18.97 & 0.620 & 0.535    \\
Ours & 20.25 & 0.705 & 0.319  & 20.70 & 0.794 & 0.272    \\
\bottomrule
\end{tabular}
}
\label{tab:UrbanScene3D}
\end{table}

\begin{figure*}[h]
    \centering
    \begin{subfigure}[t]{0.45\linewidth}
        \includegraphics[width=0.9\linewidth]{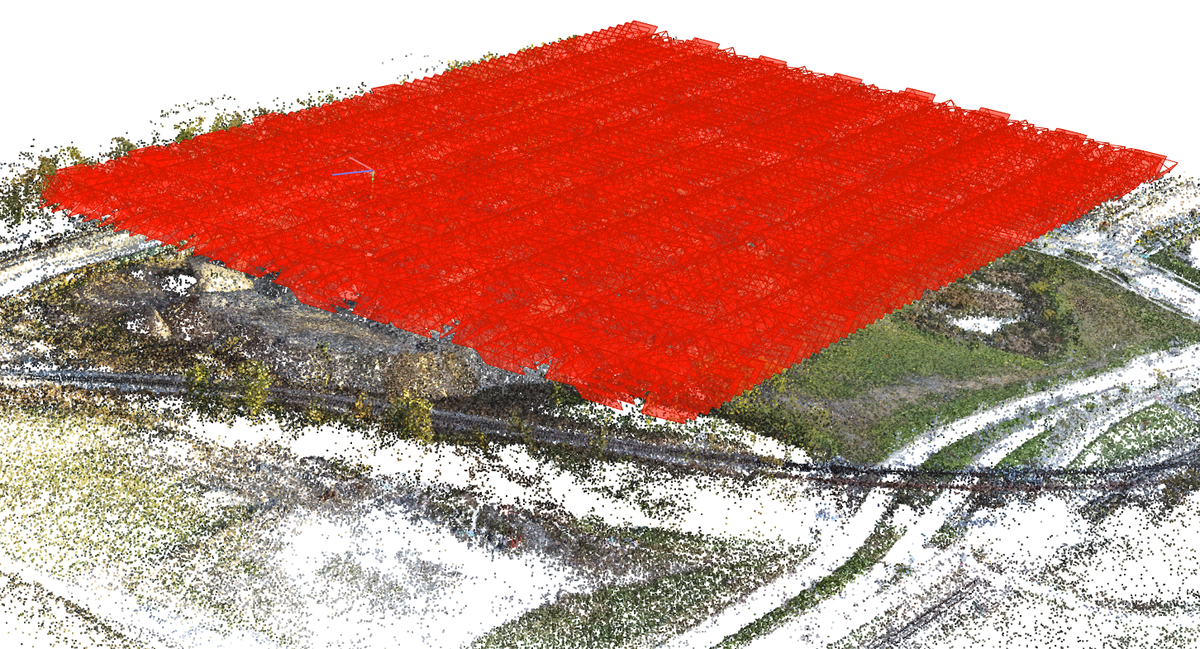}  
        \caption{Rubble (Mill19)}
        \label{rubble}
    \end{subfigure}
    \begin{subfigure}[t]{0.45\linewidth}
        \includegraphics[width=0.9\linewidth]{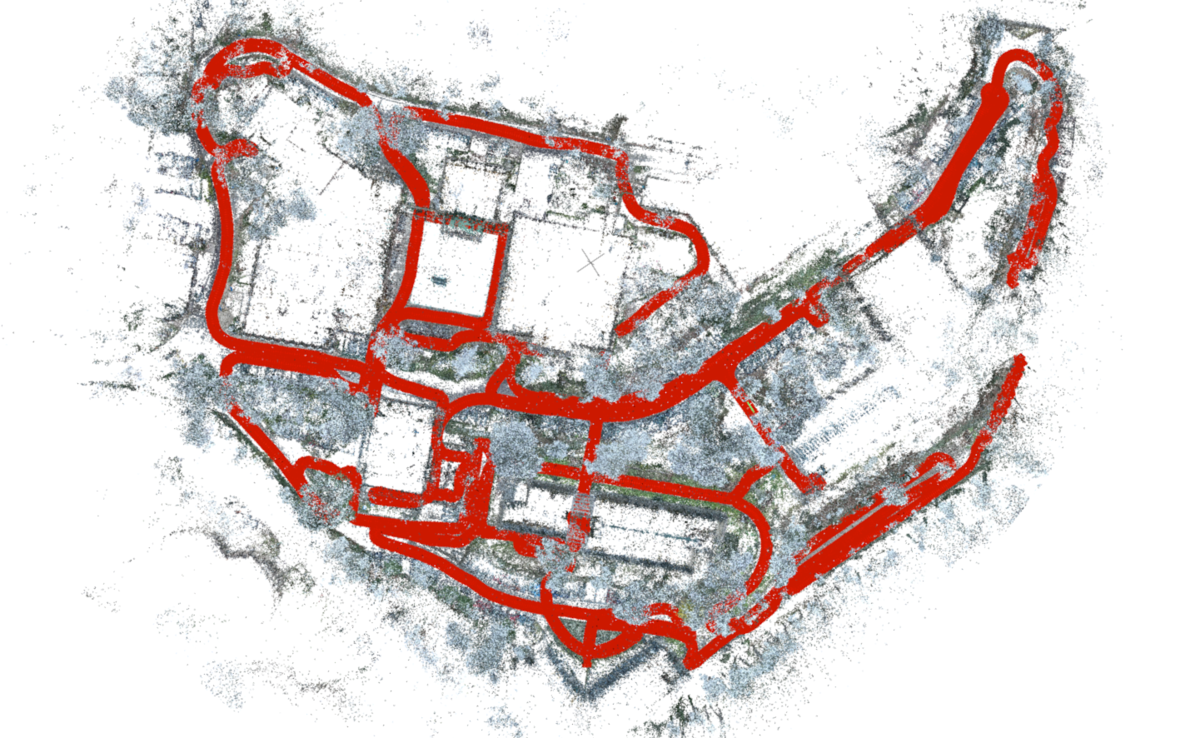}
        \caption{Campus (H-3DGS)}
        \label{campus}
    \end{subfigure}
    \caption{Overview of the aerial and street datasets.}
    \label{fig:AerialStreet}
\end{figure*}

\subsection{Evaluation Details}
\label{App:Hyperparameters}
Finding baselines for the MC-small-city+ scene proved challenging, as the only methods we are aware of that have successfully trained on all fused aerial and street-level views--\citet{RetinaGS} and \citet{NerfXL}--both require 64 GPUs running in parallel.
We were not able to meet these hardware requirements for evaluation and hold that these methods and similar multi-GPU works like \citet{Grendel} are orthogonal to our method.
H-3DGS~\citep{Hierarchical} has demonstrated the ability to reconstruct similar sized datasets, but its evaluation was restricted to street-level views.
We choose \citet{Hierarchical} and other popular large-scale Gaussian Splatting methods \citep{CityGaussian, HorizonGS, OctreeGS, CLM} as baselines. Table \ref{tab:features} compares the supported features of our method and all baselines. 

Most of the large-scale 3DGS frameworks have complicated multi-stage training processes and are very sensitive to hyperparameters, which we detail in the following section.
We consider it a benefit of our training pipeline that the user is not confronted with tuning the error-prone partitioning process, being more similar to the original Gaussian Splatting training. 

\begin{table*}[]
\begin{center}
\begin{tabular}{lcccccc}
\toprule
             & \multirow{2}{*}{\shortstack{LoD\\Training}}
             & \multirow{2}{*}{\shortstack{LoD\\Rendering}}
             & \multirow{2}{*}{\shortstack{Out-of-Core\\Training}}
             & \multirow{2}{*}{\shortstack{Interactive\\Rendering}}
             & \multirow{2}{*}{\shortstack{No Chunking\\Artifacts}}
             & \multirow{2}{*}{\shortstack{Gaussians}} \\ \\\midrule
CityGaussian &      \xmark  &    \cmark     &         \xmark       &        \cmark         &         \xmark        & Explicit  \\
HorizonGS    &      \cmark  &    \cmark     &         \xmark       &        \xmark         &         \xmark        & Neural    \\
H-3DGS       &      \xmark  &    \cmark     &         \xmark       &        \cmark         &         \xmark        & Explicit  \\
OctreeGS     &      \cmark  &    \cmark     &         \xmark       &        \cmark         &         \cmark        & Neural    \\
CLM-GS       &      \xmark  &    \xmark     &         \cmark       &        \xmark         &         \cmark        & Explicit  \\
VastGaussian &      \xmark  &    \xmark     &         \xmark       &        \cmark         &         \xmark        & Explicit  \\
OccluGaussian&      \xmark  &    \xmark     &         \xmark       &        \cmark         &         \xmark        & Explicit  \\
\midrule 
\textbf{Ours}         &      \cmark  &    \cmark     &         \cmark       &        \cmark         &         \cmark        & Explicit  \\ \bottomrule
\end{tabular}
\end{center}
\caption{Comparison of features.}
\label{tab:features}
\end{table*}

\paragraph{OctreeGS} For \emph{OctreeGS} \citep{OctreeGS}, we follow the provided instructions on training custom datasets. We trained using the suggested hyperparameters for standard scenes and those used for the MatrixCity dataset. The latter achieved higher quality metrics, which is what we report. Note that OctreeGS does not perform any scene division and uses the memory-intensive ScaffoldGS~\citep{ScaffoldGS} method, making it unsuitable for ultra-large scale scenes.

\paragraph{CityGaussian} We choose \emph{CityGaussian} \citep{CityGaussian} (specifically the 1.2 version of the repository) as a baseline instead of \emph{CityGaussianV2} \citep{CityGaussianV2}, as the code release for \emph{CityGaussianV2} does not yet support level-of-detail rendering. We follow the instructions on training large datasets and use the parameters and chunk split from the provided configuration file for the MatrixCity-Aerial dataset (which covers the same region as the MC-small-city+ dataset, but without the street views), as they produced better metrics than the default parameters. Note that the low number of Gaussians in the final model is due to \emph{CityGS} discarding most (about 90\%) of the trained Gaussians after chunk training in order to avoid artifacts. For Campus and Small City, we used the same parameters, but a $4\times4$ and $2\times2$ chunk split to match the number of chunks as closely as possible with the $12$ and $4$ chunks respectively used by \emph{H-3DGS}~\citep{Hierarchical} and a $3\times3$ chunk split for the Uni10k dataset.

\paragraph{CLM-GS}
Significant modifications were necessary to evaluate CLM-GS \citep{CLM} on our large-scale datasets. In the paper, the issue of large-scale densification is sidestepped, by starting from a dense (102 million point) ground truth point cloud and disabling densification all-together for the aerial matrix-city scene. All other evaluated scenes are sufficiently small that densification is not an issue. Running the unmodified method on large-scale datasets from sparse point clouds leads to either collapsing training or pruning of all Gaussians. Thus, we disable pruning to avoid large parts of the scene disappearing and disable opacity resets, which make the training process unstable for large-scale scenes. Experimentally we found that the high densification parameters for the 28M mode of Rubble produced the best results and used it in our evaluation. We further implemented the suggestion of \citet{Hierarchical} to densify according to maximal instead of average screen-space gradients, but did not notice significant improvements.

As suggested by the authors for large datasets, we use the clm-offload strategy with a batch size of 4, and match the number of training iterations with our method for each experiment. We enable the option to load images from disk in order to avoid going out of memory. 
As of writing the CLM codebase contains no possibility of rendering test images or evaluating performance metrics, so we implemented the necessary functions ourselves. 

\paragraph{Hierarchical-3DGS} We follow the instructions of \citet{Hierarchical} for running the method on large scenes. Note that we do not make use of their particular COLMAP pipeline for MC-smaller-city+, as it only works with unbroken camera paths.
We disable exposure optimization for evaluation in accordance with the instructions. The partitioning step of H-3DGS requires point correspondences not provided by the dataset. To substitute, we use the camera poses provided by the dataset and generate the sparse point cloud and correspondences from scratch using COLMAP. On scenes of this scale, COLMAP output contains significant noise, which also leads to reduced quality metrics for our method. \\
H-3DGS achieves drastically worse PSNR and SSIM results on the full Campus scene, compared to the single-chunk results, due to a slight perspective distortion that occurs specifically on this dataset (it did not occur on Small City).
This can be reproduced by evaluating the trained model on their Campus scene, both of which are available on their website. As such, we include the results, but encourage visual comparison in Figure 12 of the main paper.  

\paragraph{HorizonGS} We expand the configuration provided for a single chunk of the MatrixCity small-scene dataset (named \\ \textit{ours/large\_scene/block\_A}) with a chunk split of $3\times4$ and $5 \times 5$ for MC-smaller-city+ and MC-small-city+, respectively, and $2 \times 2$ for Uni10k.
Since the method throws an error if it is not given both a set of street- and aerial images, we limit evaluation to those two scenes. 
Each chunk is trained for 60000 iterations and then fine-tuned for another 40000. 

The chunk partitions are shown in Figure \ref{fig:HorizonGS_partition}; as we can see, the complexity of these partitions is vastly different.
For Uni10k, the simpler chunk partitioning leads to more stable training, and, as can be seen in Table 1, improved results.

\begin{figure}[h]
    \centering
    \includegraphics[width=0.45\linewidth]{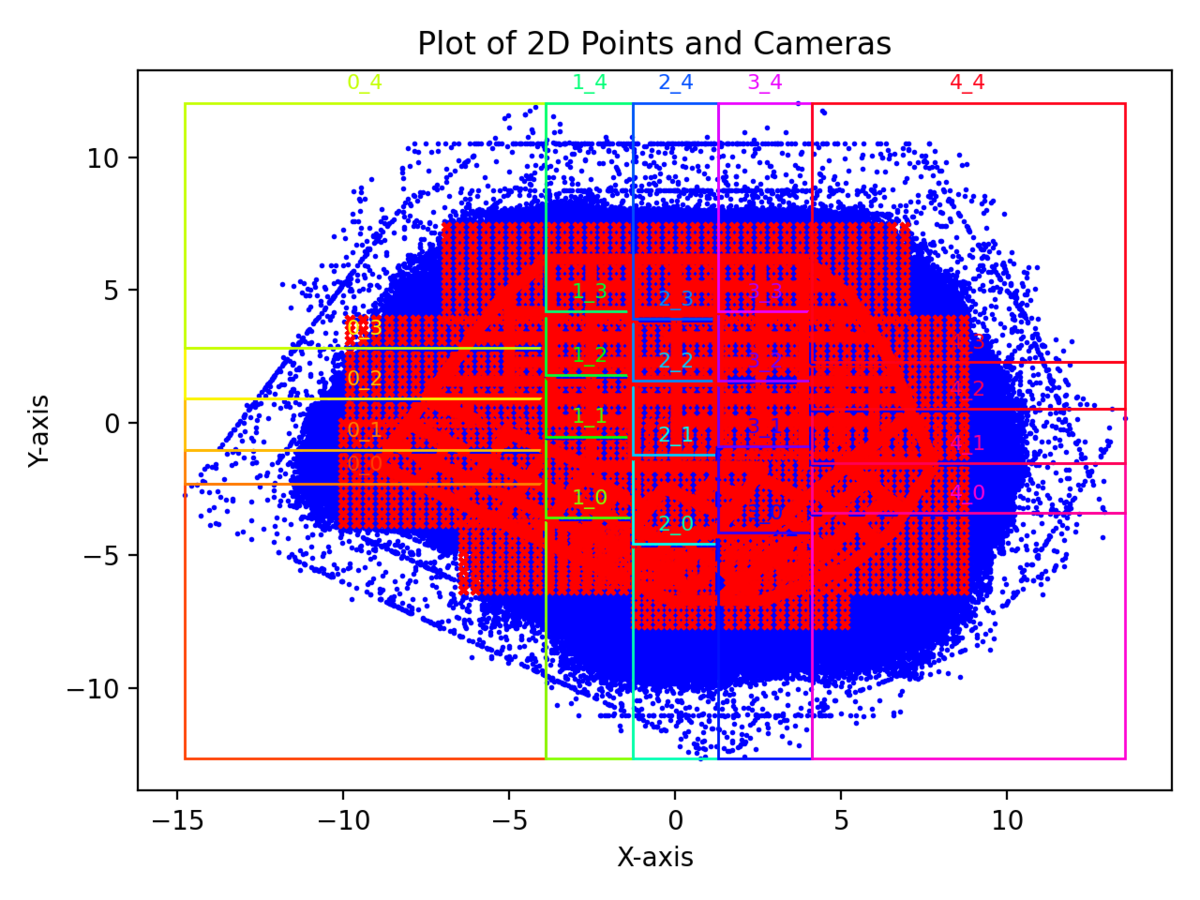}
    \includegraphics[width=0.45\linewidth]{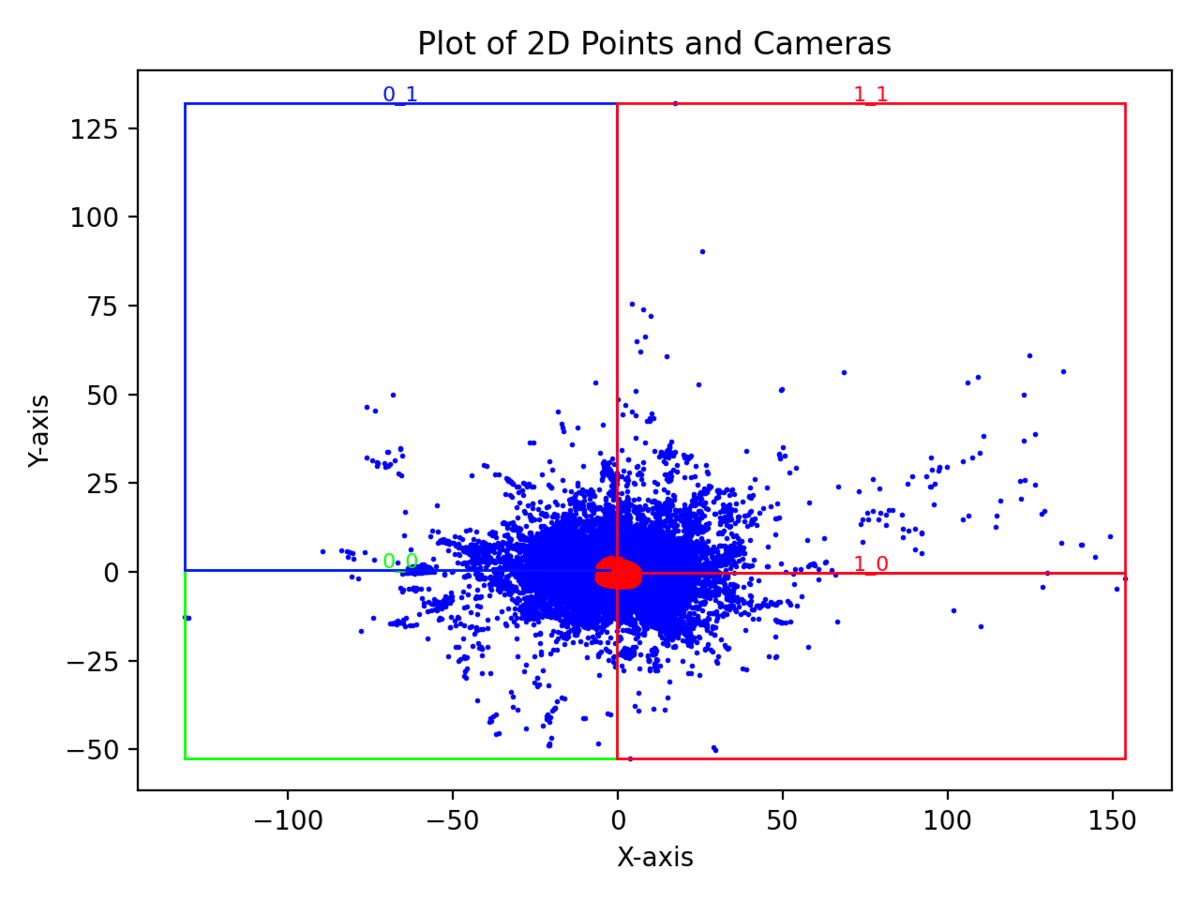}
    \caption{Partitions of HorizonGS on MC-small-city+ (left) and Uni10k (right).}
    \label{fig:HorizonGS_partition}
\end{figure}
\paragraph{VastGaussian} \emph{VastGaussian} \citep{VastGaussian} would have presented a natural comparison point, but there is currently no official code release available.

\paragraph{OccluGaussian} As of writing, the code for \citep{OccluGaussian} has not been released, so we are only able to compare to their reported results on their own dataset. Because the method is specifically tailored to indoor scenes, they never published results on any outdoor street-level datasets. 

\paragraph{Ours} In general, we perform 60k iterations of coarse training (except for 100k on MC-small-city+) and then 150k iterations of fine training (except 100k on Uni10k, 250k on Campus and MC-smaller-city+ and 500k on MC-small-city+). We choose SH degree 1 and halve the densification gradient threshold as opposed to \emph{H-3DGS}. We pick a cache size of 15 million Gaussians. For each scene, we pick a maximal number of Gaussians and desired SPT volume according to the scene scale. We deactivate the use of pinned CPU memory, which would accelerate memory transfers, because it would not be available in sufficient quantities on the consumer devices we are targeting. We provide detailed configuration files containing all hyperparameters for all of the scenes as supplemental material.



\subsection{Implementation Details}
Training models on this unprecedented scale on consumer hardware required solving many technical issues. In this section, we elaborate on the most important of these design decisions.
\paragraph{Codebase} Our general code structure is based on H-3DGS \citep{Hierarchical}.
In particular, we reuse the hierarchy creator, but replace the cut procedure and chunk based training, as well as the rasterizer for the one used in \citet{3DGS}. We also use the densification implementation of \citet{3DGSasMCMC}. All of the code is implemented in PyTorch and C++/CUDA. The entire source code is included as supplemental material.
\paragraph{Varying level of detail}
We introduce noise into the hierarchy cuts in order to prevent overfitting of Gaussians of a certain scale to views from a particular distance. In particular, we multiply the distances to the SPT centers each iteration with $1+5r^4$, with a uniformly random variable $r\in U(0,1)$.
This factor is designed such that most iterations will train near the highest level of detail, but coarser levels will also be trained occasionally. While this does not improve image metrics on the test set, it shows significant improvement to out-of-distribution views and reduces LoD popping artifacts.
\paragraph{Varying focal length}
The required level of detail is not only dependent on a camera's distance $d$ to the Gaussian, but also on its focal length, which we need to account for as our datasets contain views with differing focal lengths.
Therefore, we choose a base focal length $f_b$ and use a relative distance metric $\hat{d}$ for our HSPT cuts, which is calculated for the current camera's focal length $f_i$ as $\hat{d}=\frac{f_b}{f_i} d$.
For example, given a camera with double the focal length, the cut distance should be halved to account for the larger projection footprint on the image plane.
\paragraph{SH Degree} For the most part, experiments indicate that SH degrees higher than 1 do not significantly contribute to image quality in the tested scenes.
For training higher degrees $n$ of SH, we have found that increasing the degree from 0 to 1 during the course stage and then gradually increasing the degree from 1 to $n$ during fine training (with each increase happening after 10\% of total training iterations) yields the best results.
\paragraph{Gaussian Order}
To exploit spatial coherency and improve training performance, we store the Gaussians on CPU in Morton Z-order.
As this order can change during training, we resort the Gaussians at every densification iteration. 
\paragraph{Training Images}
Conventional Gaussian Splatting stores the entire training dataset in VRAM to achieve their impressive training speed, however, this approach is infeasible for training ultra-large-scale datasets on consumer-grade hardware.
Instead, we load the ground truth images from disk every iteration, saving VRAM at the cost of training performance.
\paragraph{Unreachable Gaussians}
During training, it can occur that a child Gaussian becomes larger than its parent.
This can lead to cases where $m_d(\text{parent}(i)) < m_d(i)$, making it impossible for the parent Gaussian to fulfil the SPT cut condition.
While we are aware of this inefficiency, we found that in practice this only happens to a small number of Gaussians ($<10\%$).
They still occupy a portion of RAM, but are never rendered or transferred. We have experimented with rebalancing the hierarchy during training, but found it too costly for little benefit. In general, we find this behaviour to be preferable over generating improper cuts, which can derail training.
\paragraph{Respawning, Densifying, and Pruning Gaussians}
For improved performance, all respawn- and densify-operations are performed in parallel during the densification step. This can lead to difficult edge cases that need to be handled to prevent the hierarchy from degenerating: When two sibling nodes both need to be respawned, we only respawn the right node. This will cause the left node to become a new dead leaf, which will be respawned in the next densification iteration. If a node needs to be respawned whose sibling is not a leaf node, its entire subtree will replace the parent node. \\ 
To minimize the number of unnecessary Gaussians, we apply a simple pruning strategy where we zero the opacity of Gaussians that have not contributed for a number of iterations equal to twice the size of the training set.
This will cause them to be respawned in the next densification iteration.
\paragraph{Performance Optimization}
The performance of the HSPT cut and Gaussian loading is particularly important, as they happen every iteration. We store the SPT properties ($m_d(i)$, $m_d(\text{parent}(i))$, $i$) for all Gaussians in a single continuous GPU memory buffer and perform the cuts in parallel using optimized CUDA kernels.
Similarly, the Gaussian properties are stored in a single PyTorch tensor in RAM as an array of structures (each Gaussian's properties concurrently).
The properties of all required Gaussians are then transferred to the GPU via a single copy operation, reorganized on the GPU in a structure of arrays layout (as required by the rasterizer), and appended to the current render set.

\subsection{Additional Densification Details}
\label{App:Densification}
Our densification method combines the split operation from \citet{3DGSasMCMC} and the selection method from \citet{Hierarchical} with coarse-to-fine expansion of the LoD structure.
Relying purely on the densification strategy of \citet{3DGSasMCMC}--applying a noise to every Gaussian's position and an opacity/scaling loss--becomes problematic for large scenes, where a majority of Gaussians are not visible from any one view: 
The result is a gradual disappearance of density in the scene or extreme ''stringing'' artifacts (cf. Figure \ref{fig:MCMCFail}).
However, without these losses to encourage respawning of Gaussians, the densification becomes aimless and random.
We implemented another strategy that only applies the losses and noise to Gaussians that affected at least one pixel in the current iteration. While this significantly improved results, we found that the MCMC densification strategy in general performs poorly when Gaussian density is low, which is necessitated by ultra-large-scale scenes. On small scenes such as the single-chunk datasets, our modified MCMC strategy slightly outperforms the presented strategy given sufficient Gaussian budget, but it lags behind in the larger scenes. Avoiding destructive opacity resets, as advocated by \citet{3DGS}, is particularly important in our setting, as many Gaussians will no longer conform to the hierarchical structure after opacity is restored to normal.
\begin{figure}[h]
    \centering
    \begin{subfigure}{0.45\linewidth}
        \includegraphics[width=0.9\linewidth]{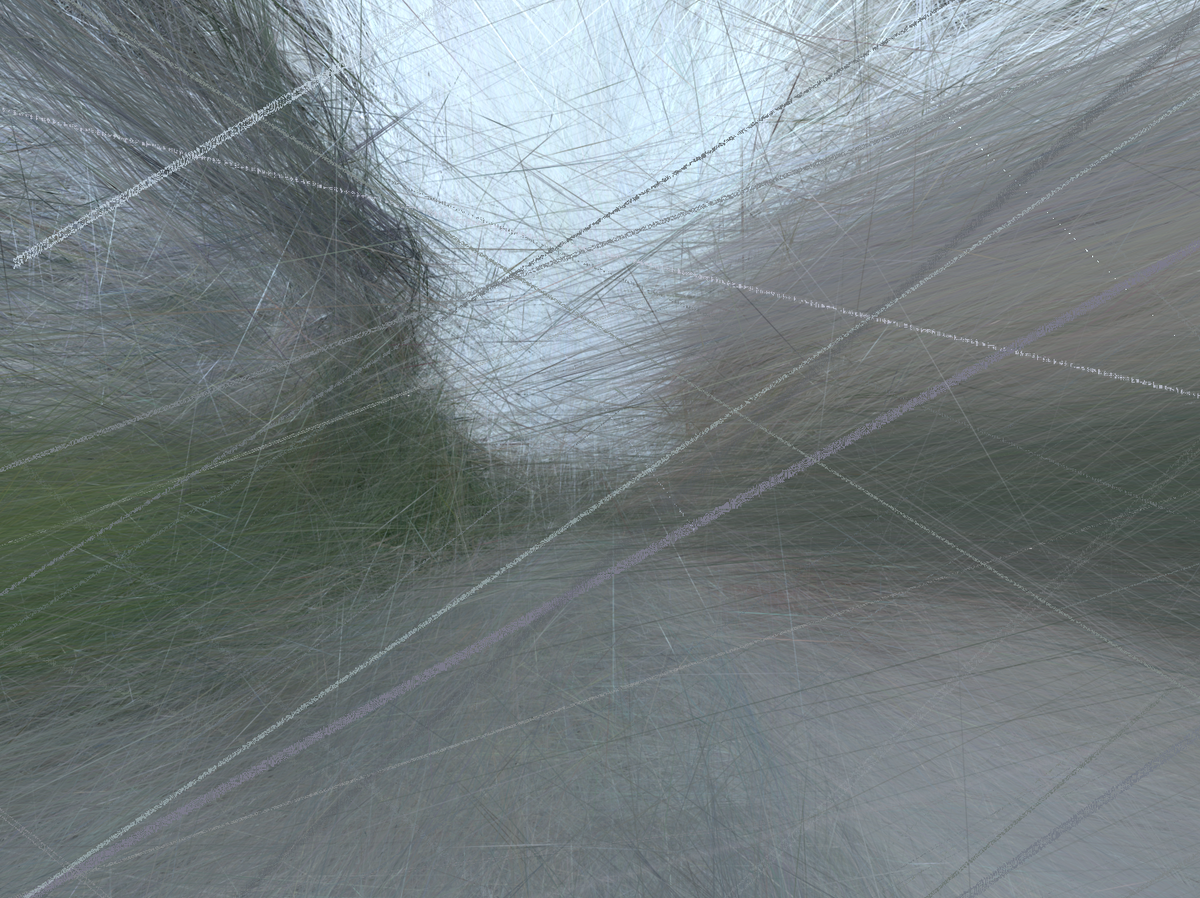}
        \caption{Campus}
    \end{subfigure}
    \begin{subfigure}{0.45\linewidth}
        \includegraphics[width=\linewidth]{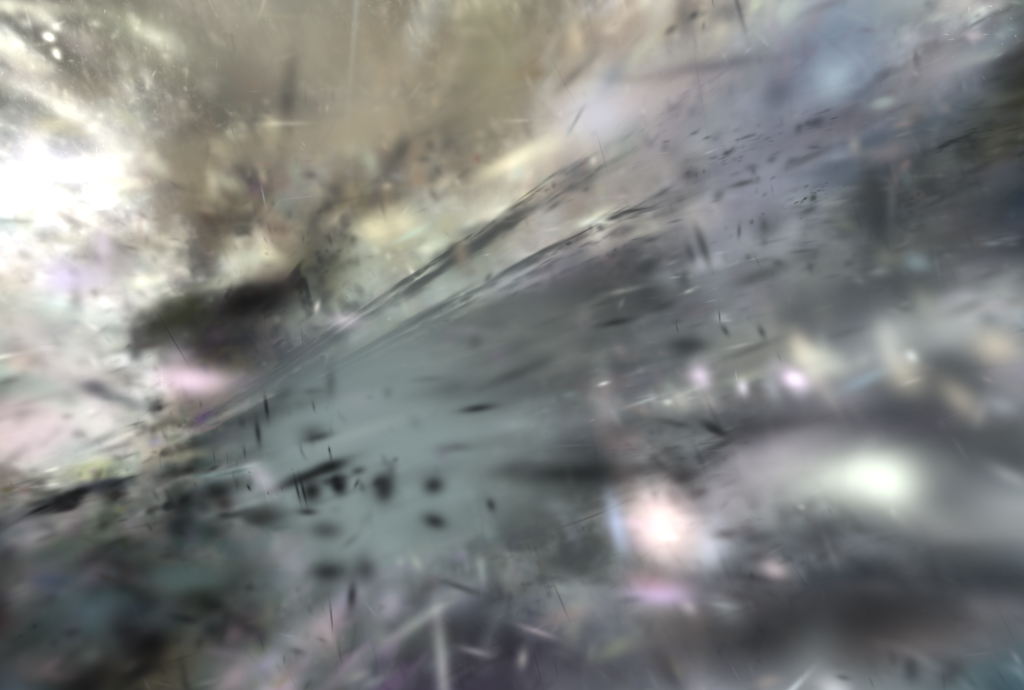}
        \caption{Small City}
    \end{subfigure}
    \caption{Without regularization, the scale- and opacity-losses of 3DGS-as-MCMC lead to either stringing or disappearance of the entire scene.}
    \label{fig:MCMCFail}
\end{figure}

\subsection{Pseudocode}
\label{App:Pseudocode}
This section provides pseudocode for our core algorithms.
Algorithm \ref{alg:BFScut} contains a procedure to ''cut'' a binary tree hierarchy along a certain condition.
Algorithm \ref{alg:SPTcut} shows how to cut a single SPT at a particular distance $d^j$.
The entire training procedure is sketched in Algorithm \ref{alg:cut}.

\begin{algorithm}[h]
\caption{BFS Hierarchy Cut}\label{alg:BFScut}
\begin{algorithmic}
\Procedure{Cut}{Hierarchy $\mathcal{H}, condition$}
\State $cut \gets \{\}$
\State $Q \gets \textsc{Queue}(root(\mathcal{H}))$
\While{$Q$ not empty}
\State $node \gets Q.\textsc{dequeue}()$
\If {$condition(node)$}
\State $cut \gets cut \cup node$
\Else
\State $Q.\textsc{enqueue}(node\rightarrow left)$
\State {$Q.\textsc{enqueue}(node\rightarrow right)$}
\EndIf
\EndWhile{}
\State $\texttt{return } cut$
\EndProcedure
\end{algorithmic}
\end{algorithm}

\begin{algorithm}[h]
\caption{SPT Cut}\label{alg:SPTcut}
\begin{algorithmic}
\Procedure{Cut}{SPT $\mathcal{S}$, distance $d^j$}
\State $high \gets \textsc{BinarySearch}(\mathcal{S}_{max},d^j)$
\State $cut \gets \{i \in [0, high] | \mathcal{S}_{min}^i < d^j\}$
\State $\texttt{return } cut$
\EndProcedure
\end{algorithmic}
\end{algorithm}

\begin{algorithm*}[h]
\caption{Full train procedure}\label{alg:cut}
\begin{algorithmic}
\Procedure{Train}{upper Hierarchy $\mathcal{U}, cache, skybox$}
\State $view \gets 0$
\State $Cache\_Distances \gets \{\}$
\State $Cache\_SPTs \gets \{\}$
\While{$True$}
\State $view \gets \textsc{Sample}(knn\_graph, view)$
\Comment{Find a new nearby view}
\State $cut\_condition \gets [\lambda(i) = \neg\ is\_in\_frustum(\mu_i) \lor m_d(i) < T ]$
\State $upper\_cut \gets \textsc{cut}(\mathcal{U}, cut\_condition)$
\State $SPT\_roots \gets \{u \in upper\_cut | u \text{ contains an SPT}\}$
\State $upper\_nodes\_to\_render \gets upper\_cut \setminus SPT\_roots$
\State $SPT\_distances \gets \{\left\| \mu_r - p_{view}\right\|_2^2 \text{ for } r\in SPT\_roots\}$
\State $Cache\_Reuse\_SPTs \gets \{s_i \in (Cache\_SPTs \cap SPT\_roots) | D_{min}\leq \left\| \frac{\mu_i - \mathbf{p}_{cam}}{Cache\_Distances_i}\right\|_2^2 \leq D_{max} \}$
\State $load\_SPTs \gets (SPT\_{roots} \setminus Cache\_Reuse\_SPTs)$
\State $ load\_Gaussian\_indices \gets \textsc{Cut\_SPTs}(load\_SPTs, SPT\_distances[load\_SPTs])$
\State $load\_Gaussians \gets \textsc{Load}(load\_Gaussian\_indices)$
\State $Gaussians \gets skybox \cup upper\_nodes\_to\_render \cup load\_Gaussians$
\State $Gaussians \gets Gaussians \cup \textsc{load\_cache}(Cache\_Reuse\_SPTs)$
\State $\textsc{Render\_And\_Optimize}(Gaussians)$
\State $Cache\_SPTs \gets Cache\_SPTs \cup load\_SPTs$
\If {$len(Cache\_Gaussians) > max\_cache\_size$}
\State $\textsc{Write\_Back\_LRU}(Cache\_Gaussians)$
\Comment{Write cached Gaussians to RAM}
\EndIf
\EndWhile
\EndProcedure
\end{algorithmic}
\end{algorithm*}


\end{document}